\documentclass[twocolumn,aps,prl,superscriptaddress]{revtex4-2}

\usepackage{amsmath}
\usepackage{graphicx}
\usepackage{MnSymbol}
\usepackage{parskip}
\usepackage{makecell}
\usepackage{booktabs}
\usepackage{multirow}
\usepackage{xcolor}

\begin{document}

\title{Adhesion-controlled sliding and the Stribeck curve in hydrophobic soft contacts}

\author{R. Xu}
\affiliation{Peter Gr\"unberg Institute (PGI-1), Forschungszentrum J\"ulich, 52425, J\"ulich, Germany}
\affiliation{State Key Laboratory of Solid Lubrication, Lanzhou Institute of Chemical Physics, Chinese Academy of Sciences, 730000 Lanzhou, China}
\affiliation{MultiscaleConsulting, Wolfshovener str. 2, 52428 J\"ulich, Germany}

\author{C. Spies}
\affiliation{Robert Bosch GmbH, Corporate Research and Advance Engineering, Robert-Bosch-Campus 1, 71272 Renningen, Germany}
\affiliation{University of Freiburg, Department of Microsystems Engineering, Georges-K\"ohler-Allee 103, 79100 Freiburg, Germany}

\author{M. Scaraggi}
\affiliation{Department of Engineering for Innovation, University of Salento, 73100, Lecce, Italy}
\affiliation{Center of Biomolecular Nanotechnologies, Istituto Italiano di Tecnologia, 73010, Arnesano, Italy}
\affiliation{State Key Laboratory of Solid Lubrication, Lanzhou Institute of Chemical Physics, Chinese Academy of Sciences, 730000 Lanzhou, China}

\author{B.N.J. Persson}
\affiliation{Peter Gr\"unberg Institute (PGI-1), Forschungszentrum J\"ulich, 52425, J\"ulich, Germany}
\affiliation{State Key Laboratory of Solid Lubrication, Lanzhou Institute of Chemical Physics, Chinese Academy of Sciences, 730000 Lanzhou, China}
\affiliation{MultiscaleConsulting, Wolfshovener str. 2, 52428 J\"ulich, Germany}

\begin{abstract}
We present an experimental and theoretical study of dry and glycerol-lubricated sliding for polymethyl methacrylate (PMMA) cylinders with different surface roughness sliding on polydimethylsiloxane (PDMS) rubber. This system represents a hydrophobic soft contact, where adhesion may persist even in the presence of the lubricant and thereby modify both the real contact area and the sliding response. Dry-friction measurements, combined with contact-area calculations that include adhesion, provide a baseline for the lubricated study. For the two sandblasted surfaces, the measured Stribeck curves are described reasonably well by a mean-field mixed-lubrication theory with a fitted velocity-independent effective interfacial shear stress. In contrast, the smooth surface exhibits qualitatively different behavior. 
We attribute this to an adhesion-controlled sliding mode involving macroscopic Schallamach-wave-like instabilities at low sliding speeds, which are progressively suppressed as the sliding speed increases and forced wetting reduces direct solid-solid contact. The results show that, for soft hydrophobic contacts, the Stribeck curve cannot always be understood from classical fluid flow and load sharing alone. For sufficiently smooth and adhesive surfaces, adhesion changes not only the real contact area but also the sliding mode itself.
\end{abstract}

\maketitle

{\bf Corresponding author:} B.N.J. Persson, email: b.persson@fz-juelich.de

\setcounter{page}{1}
\pagenumbering{arabic}




\section{Introduction}

Understanding the nature of fluid flow at sliding interfaces is one of the central topics in tribology
because of its practical importance \cite{Huge1,Huge2,Huge3,Jagota,Jagota1}. Of fundamental importance is the
Stribeck curve, which shows how the friction in fluid-lubricated contacts depends
on the sliding speed. The Stribeck curve depends nonlinearly on the contact load,
the lubricant viscosity, and the lubricant entrainment speed. Of particular importance
is the transition from fluid lubrication, which occurs at sufficiently high sliding speed
when the local surface separation is everywhere larger than the amplitude
of the surface roughness, to mixed lubrication and boundary lubrication,
where direct solid-solid contact occurs.

The Stribeck curve for elastically stiff materials like steel at low sliding speed may involve
asperity contact regions where the lubricant fluid is confined at very high pressures, making the
pressure dependence of the fluid viscosity important.
This is a very complex and not fully understood topic \cite{Huge3}.
Here we are interested in contacts where at least one of the solids is
elastically soft \cite{PS1,PS2,PS3,PS4,PS5,Huge1,last}. In this case,
the pressure dependence of the fluid viscosity is unimportant. However, the shear rate may still be high, and the dependence of the viscosity on the shear rate may be relevant and typically results in shear thinning \cite{thin}. Under such conditions, thermal effects may also become important, and the combined effect of shear thinning and thermal effects may contribute to the friction response \cite{Rx}. For molecularly thin films between locally smooth surfaces,
the lubricants may also form layers parallel to the surfaces \cite{Israel},
which influence the shear properties of the fluids \cite{Persson}. For very smooth surfaces
and for lubricants interacting very weakly with the solid walls, wall slip may occur \cite{Persson}, which can have a large influence on the friction force \cite{Paolo}.

In this paper we present a detailed experimental and theoretical study of the Stribeck curves for polymethyl methacrylate (PMMA)
cylinders with different surface roughness sliding on flat polydimethylsiloxane (PDMS) rubber surfaces lubricated by glycerol.
For this system we find that treating glycerol as a Newtonian fluid results in good agreement between theory and experiment for the sandblasted surfaces.
The main aim of this study is to examine to what extent the mixed-lubrication theory developed in Ref. \cite{PS3} can describe the Stribeck curves of hydrophobic soft contacts with different surface roughness. In addition, we present results for the sliding friction of dry PMMA-PDMS contacts. 
These dry-friction studies are important because they provide a baseline for the interfacial shear mechanism and for the interpretation of the lubricated friction data.

One ``complication'' with the PMMA-glycerol-PDMS system is that for stationary contact, the free energy is reduced
when glycerol is removed from the solid contact area. We refer to this case as a hydrophobic interface with respect
to glycerol. In this case, adhesion occurs between the contacting solid walls even when the solids are immersed in glycerol.
This results in an adhesion-induced increase in the real contact area, which is not included in any elastohydrodynamic
computer simulation software we are aware of. Thus, for this type of system, the central question is whether the Stribeck curve can still be understood within the standard mixed-lubrication framework, or whether adhesion changes not only the real contact area but also the sliding mechanism in a more fundamental way. In most practical applications, for lubricated interfaces, the fluid wets the solid
walls, and a thin fluid film separates the surfaces unless the squeezing pressure is high enough. For such wetting systems,
there is no adhesion between the solids in the fluid, and the contact mechanics is accurately described using, e.g., the Persson
contact mechanics theory. However, here we focus mainly on the physically more interesting case of hydrophobic interfaces.

The paper is organized as follows. In Sec. 2, we present the basic equations used when calculating Stribeck curves.
Sec. 3 describes the different experimental setups used to study friction,
adhesion, and surface topography. In Sec. 4, we present the power spectra of all surfaces and the calculated relative contact area
as a function of the nominal contact pressure. Sec. 5 presents the results of the adhesion study
for both dry and lubricated (with glycerol) contact between the silicone rubber and PMMA.
In Sec. 6, we show the dependence of the friction coefficient
on the sliding speed for dry contacts. In Sec. 7, we present the measured data for the Stribeck curve, which we compare with the theoretical
predictions. The experimental and theoretical results are discussed in Sec. 8,
and the summary and conclusions are given in Sec. 9. In Appendix A, we show the fluid flow and friction
factors used in the calculations of the Stribeck curve, and in Appendix B, we show
the fluid and contact pressure distributions and the surface separation as a function of the lateral coordinate $x$.


\begin{figure}
\includegraphics[width=0.35\textwidth,angle=0.0]{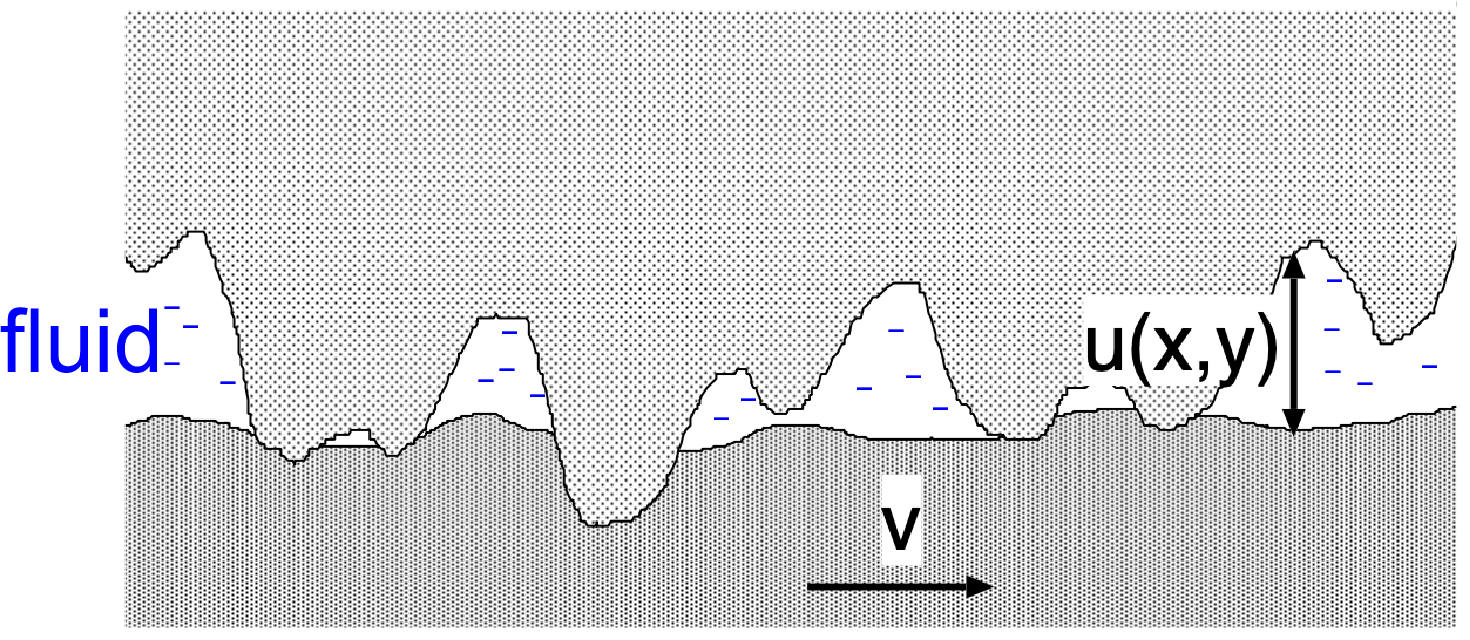}
\caption{\label{ContactGeometry.eps}
A rigid body is squeezed into contact with an elastic solid with a flat surface in a fluid.
In the actual setup, the rough surface (upper) is stationary, and the substrate (lower) moves with speed $v$.
}
\end{figure}

\vskip 0.3cm
\section{Mean-field theory for mixed lubrication in rough soft contacts}

The experimental sliding friction data will be analyzed using the theory developed in Ref. \cite{PS3},
and here we review some of the basic equations. We consider a rigid body with surface roughness (PMMA cylinders)
sliding on an elastic substrate (PDMS rubber) with a flat and smooth surface, lubricated by a Newtonian fluid (glycerol) with viscosity
$\eta$, as shown in Fig. \ref{ContactGeometry.eps}. The ensemble-averaged flow current can be written as
$$\bar {\bf J} = - { \bar u^3\over 12 \eta}  \phi_{\rm p} (\bar u) \nabla \bar p_{\rm fluid} +
{1\over 2} \bar u {\bf v} +{1\over 2} h_{\rm rms} \phi_{\rm s} (\bar u) {\bf v} \eqno(1)$$
Here $\bar p_{\rm fluid} ({\bf x}) = \langle p_{\rm fluid} ({\bf x}) \rangle$ is the ensemble-averaged fluid pressure, and similarly for the
surface separation $\bar u$ and the flow current $\bar {\bf J}$. The surface root-mean-square roughness is denoted by $h_{\rm rms}$. The fluid
pressure and shear flow factors \cite{Pati1,Pati2} 
$\phi_{\rm p} (\bar u)$ and $\phi_{\rm s} (\bar u)$ depend on the surface separation $\bar u$, and formally result
when the surface roughness is ``integrated out.'' We are interested in steady-state sliding, 
and since the fluid is assumed to be incompressible, the conservation of fluid volume gives
$$\nabla \cdot \bar {\bf J} =0 \eqno(2)$$
If we define the $x$ axis as the sliding direction and the $y$ axis as the orthogonal direction, then
for cylinders, (2) reduces to $d \bar J_x /dx=0$, or $\bar J_x$ is a constant, which
we denote by $v u^*/2$. Using this we can write (1) as
$$- {\bar u^3 \over 12 \eta} \phi_{\rm p} (\bar u) {d \bar p_{\rm fluid} \over dx} +
{1\over 2} \bar u v +{1\over 2} h_{\rm rms} \phi_{\rm s} (\bar u) v= {1\over 2} vu^* \eqno(3)$$
For surfaces without roughness, the flow factors are $\phi_{\rm p} =1$ and $\phi_{\rm s}=0$. We calculate the flow factor $\phi_{\rm p}$ using $\phi_{\rm p} = 12 \eta \sigma_{\rm eff}/\bar u^3$,
where the fluid flow conductivity $\sigma_{\rm eff}$ is obtained using the Bruggeman effective medium theory \cite{Brug}, modified to give the correct percolation condition at the relative contact area $A/A_0 \approx 0.42$, as found in accurate numerical studies for randomly rough surfaces (see Ref. \cite{MP}). For the fluid flow factor $\phi_{\rm s}$ we use a formula that
interpolates between the known results for large and small $\bar u$ (see Ref. \cite{PS3}).

The friction coefficient is obtained by integrating the ensemble-averaged frictional shear stress, $\bar \sigma (x)$,
over the surface area and dividing by the normal force (load).
The ensemble-averaged frictional shear stress can be written as
$$\bar \sigma = (\phi_{\rm f}+\phi_{\rm fs}) {\eta v \over \bar u} + \phi_{\rm fp} {1\over 2} \bar u \nabla \bar p + {A\over A_0} \sigma_{\rm f} \eqno(4)$$
The most important friction factor is $\phi_{\rm f} = \bar u \langle u^{-1} \rangle$, which can be
obtained from the probability distribution of interfacial separation, $P(u)$, which we determine using the Persson contact mechanics theory.
In Ref. \cite{PS3}, it is shown how the other two friction factors, $\phi_{\rm fs}$ and $\phi_{\rm fp}$, can be obtained approximately.

To calculate the Stribeck curve, the averaged fluid flow equation must be coupled to the contact mechanics relation between the mean interfacial separation and the contact pressure. In the present work, the Stribeck curve is obtained by solving the fluid flow and contact mechanics equations self-consistently within a mean-field-type framework. At each position, the applied normal pressure is shared between the fluid and the solid-solid contact. Thus, the ensemble-averaged pressure $\bar p$ is written as the sum of the ensemble-averaged fluid pressure $\bar p_{\rm fluid}$ and the ensemble-averaged contact pressure $\bar p_{\rm cont}$:
$$\bar p = \bar p_{\rm fluid}+\bar p_{\rm cont}\eqno(5)$$
The separation $\bar u$ is related to the contact pressure $\bar p_{\rm cont}$ using the Persson contact mechanics theory.

The relative contact area $A/A_0$ in (4) is given by
$$ {A\over A_0} = {\rm erf} \left( {1\over 2\surd{G}} \right) \eqno(6) $$
with
$$ G =\left ({E^* \over \bar{p}_{\rm cont}}\right )^2 {\pi \over 4} \int_{q_0}^{q_1} dq \, q^3 \, C(q) \eqno(7) $$
Here, $q_0$ and $q_1$ are the long- and short-distance cutoff wavenumbers,
which in the present case are $ q_0 \sim \pi /d_0$ and $q_1 \sim \pi /d_1$, where
$d_0$ is the width of the Hertz contact region and $d_1 \sim 1 \ {\rm nm}$.
$E^*=E/(1-\nu^2)$, where $E$ and $\nu$ are the Young's modulus and Poisson's ratio of the considered material.

Using that ${\rm erf}(x) \approx 2x / \surd{\pi}$ for $x\ll 1$, equation (6)
shows that as long as $\bar p_{\rm cont}$ is small, the area of real contact is proportional 
to $\bar p_{\rm cont}$, and we have
$$\bar p_{\rm cont} \approx \beta E^* e^{-\alpha \bar u /h_{\rm rms}}\eqno(8)$$
where $\alpha$ and $\beta$ are determined by the surface roughness power
spectrum, but typically $\alpha \approx 0.5$ and $\beta \approx 0.4 q_{\rm r} h_{\rm rms}$,
where $q_{\rm r}$ is the roll-off wavenumber of the surface roughness power spectrum.
When the pressure is so high that the relative contact area is larger than $\approx 0.3$, then (8) is not valid and the relation
between $\bar p$ and $\bar u$ must be determined numerically using the equations presented in Ref. \cite{separation1,separation2,separation3}.

The pressure $\bar p (x)$ deforms the rubber surface so that for the cylinder geometry, the surface separation is
$$\bar u(x) = \bar u_0 +{x^2\over 2R} -{2\over \pi E^*} \int_{-\infty}^\infty dx' \, \bar p(x') {\rm ln} \left | {x-x'\over x'}\right | \eqno(9)$$
The constant $u_0$ is determined by the normalization condition
$$\int_{-\infty}^\infty dx \, \bar p(x) = {F_{\rm N}\over L}\eqno(10)$$

For the adhesion contribution discussed in Sec. 5, we also introduce the magnification-dependent apparent contact area.
We define $q=q_0 \zeta$, where $\zeta$ is the magnification. The apparent contact area when the interface is observed
at the magnification $\zeta$ is denoted by $A(q)$
and is obtained by replacing the upper integration limit $q_1$ in (7) with $q=q_0 \zeta$. The magnification-dependent
contact area enters the theory of adhesion, where we denote $A(q)/A_0$ by $P(q)$, which is the probability
that the surfaces make contact when the interface is observed at the magnification $\zeta$.
The area $A$ in (4) is the true contact area, i.e., $A(q_1)$, obtained when all surface roughness
components with wavenumbers between $q_0$ and $q_1$ are included.

\vskip 0.3cm
\section{Experimental setup and measurements}

We measured the friction coefficient as a function of sliding speed (Stribeck curve) for three different PMMA cylinders sliding on a flat silicone rubber (PDMS) slab, both in dry conditions and
in glycerol. The PMMA cylinders have radius $R=7 \ {\rm cm}$ and axial length $L=10 \ {\rm cm}$. The three cylinders were cut
from the same tube but were subjected to different surface treatments, resulting in different roughness levels. One surface was sandblasted for 1 min
at an air pressure of $1 \ {\rm MPa}$ and is denoted as sandblasted-rough (Sb-R), one was sandblasted for 0.5 min
at an air pressure of $0.5 \ {\rm MPa}$ and is denoted as sandblasted-smooth (Sb-S), and one was left untreated and is denoted as Smooth.

The topography of all surfaces was measured using
a Mitutoyo Portable Surface Roughness Measurement Surftest SJ-410 profilometer equipped with a diamond tip having a radius of curvature of
$2 \ {\rm \mu m}$, and with a tip-substrate loading force of $0.75 \ {\rm mN}$.
The step length (pixel size) is $0.5 \ {\rm \mu m}$, the scan length is $25 \ {\rm mm}$,
and the tip speed is $50 \ {\rm \mu m/s}$. Assuming isotropic roughness, the 2D power spectra shown below were calculated from the 1D power spectra obtained by averaging over three profile measurements in the radial (sliding) direction on each surface \cite{XuP}.

The silicone rubber was prepared from Sylgard 184, a two-component kit purchased from Dow Corning (Midland, MI), consisting of a base (vinyl-terminated polydimethylsiloxane) and a curing agent
(methylhydrosiloxane-dimethylsiloxane copolymer) with a suitable catalyst. From these two components we prepared mixtures with a 1:10 ratio
(cross-linker/base) by weight. The mixture was degassed and then cured in
an aluminum container at room temperature for 3 days.

The sliding tests were performed using a linear friction slider at four normal forces: $F_{\rm N}=31$, $91$, $151$, and $211 \ {\rm N}$. The corresponding forces per unit length are $F_{\rm N}/L = 310$, $910$, $1510$, and $2110 \ {\rm N/m}$, and, using the nominal contact area obtained from Hertzian dry line contact, the corresponding nominal contact pressures are $\sigma_0 = 0.042$, $0.072$, $0.092$, and $0.109 \ {\rm MPa}$. The aluminum box was attached to the machine table, which moved transversely using a servo drive via a gearbox. This setup allows precise control of the relative velocity between the rubber specimen and the PMMA cylinder. The sliding speed was varied from $v=1 \ {\rm \mu m/s}$ to $1 \ {\rm cm/s}$, while the force cell records the normal and friction forces. For each imposed sliding speed, the friction coefficient was extracted from the corresponding segment of the raw force signal. The detailed procedure used to calculate the reported friction coefficients and error bars is given in Appendix C.

Measurements were first performed in glycerol (purity 99.5\%) with a sufficiently large film thickness ($\sim 1 \ {\rm cm}$) to ensure fully flooded conditions. New glycerol was used for each measurement to avoid
a reduction in fluid viscosity due to absorption of water from the atmosphere. After the lubricated tests, dry tests were performed.


\begin{figure}
\includegraphics[width=0.47\textwidth,angle=0.0]{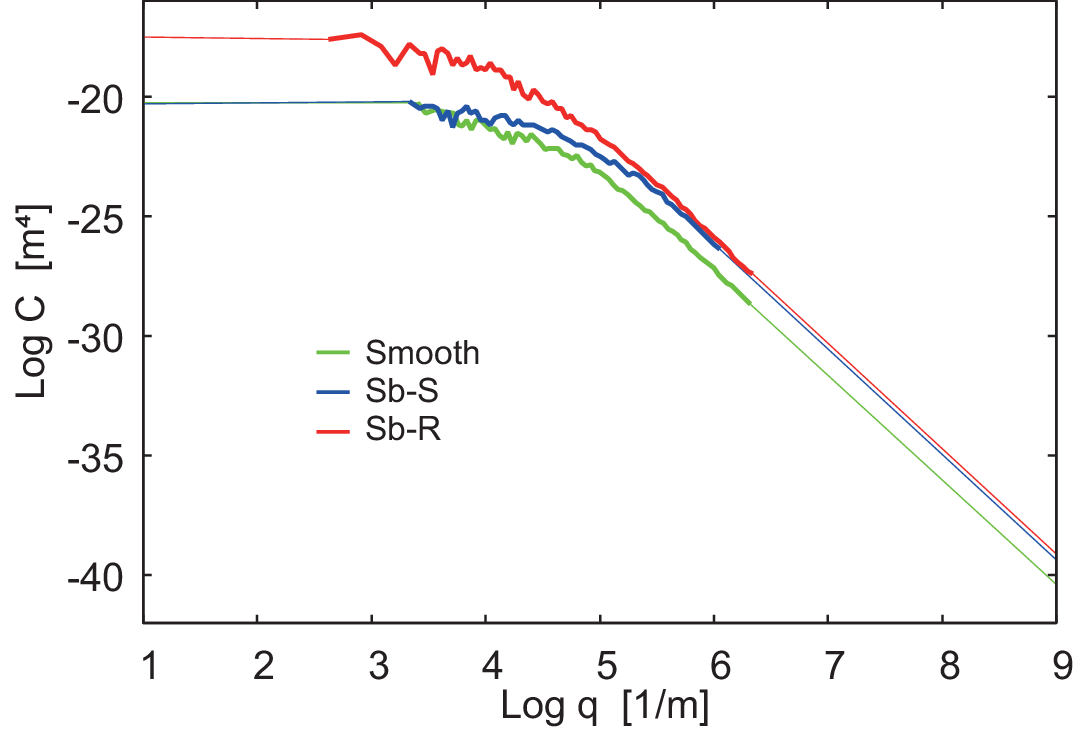}
\caption{\label{cut1.and.not.logq.2logC.all.full.eps}
The thick lines show the measured surface roughness power spectra as a function of wavenumber (log-log scale) for
Sb-S (blue line), Sb-R (red line), and the Smooth surface (green line).
The thin lines indicate the extrapolated regions at small and large wavenumbers.
}
\end{figure}

\begin{figure}
\includegraphics[width=0.47\textwidth,angle=0.0]{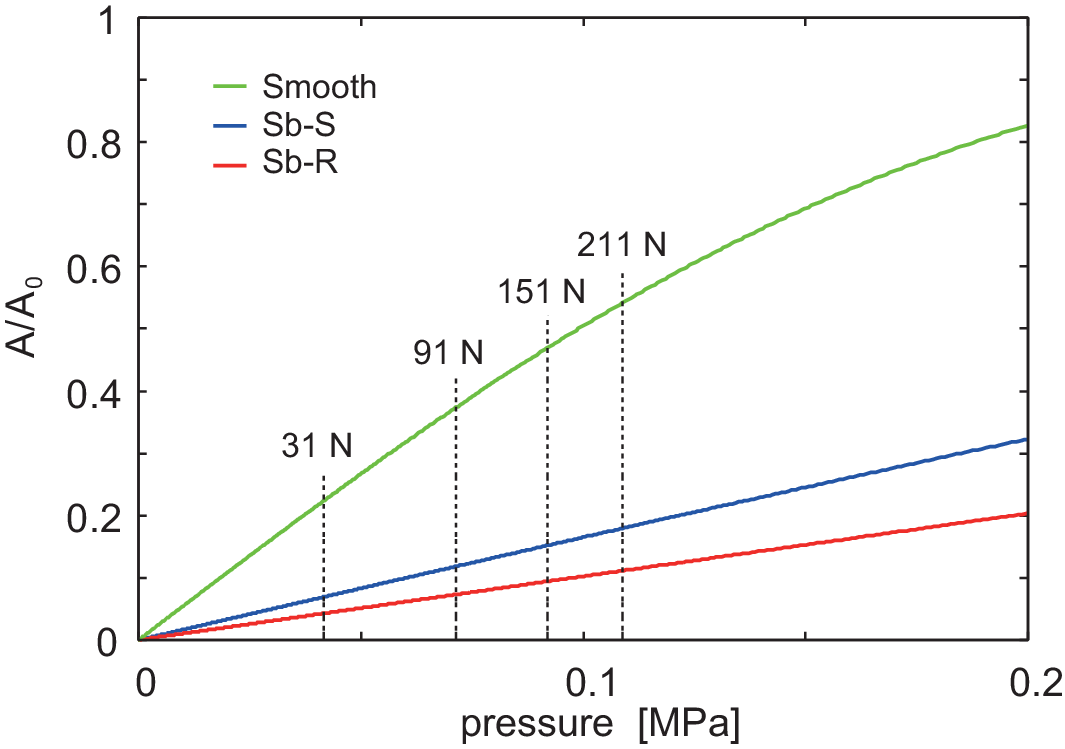}
\caption{\label{1pressure.2AreaSoothRough1.eps}
The relative contact area as a function of nominal pressure for Sb-S (blue line), Sb-R (red line), and the Smooth surface (green line).
The dotted vertical lines indicate the average Hertzian line-contact pressures (nominal contact pressures)
at the tested normal forces.
}
\end{figure}

\vskip 0.3cm
\section{Surface topography and contact area}

The thick solid lines in Fig. \ref{cut1.and.not.logq.2logC.all.full.eps}
show the 2D power spectra $C(q)$ calculated from the measured height profiles for the three PMMA cylinders: Sb-S (blue line), Sb-R (red line), and Smooth (green line).
Note that the measured power spectra are extrapolated to larger and smaller wavenumbers to cover all relevant length scales (thinner lines in Fig. \ref{cut1.and.not.logq.2logC.all.full.eps}).
Including both the measured roughness components and the extrapolated parts of the power spectra at smaller and larger wavenumbers for Sb-S, Sb-R, and Smooth gives the root-mean-square (rms) roughness amplitudes $h_{\rm rms} = 2.80$, $15.76$, and $1.37 \ {\rm \mu m}$, respectively, with corresponding rms slopes of $0.480$, $0.778$, and $0.147$.

Fig. \ref{1pressure.2AreaSoothRough1.eps} shows the
relative contact area as a function of nominal pressure for Sb-S (blue line), Sb-R (red line), and Smooth (green line). The calculations assume no adhesion
and use a Young's modulus of $E=1.5 \ {\rm MPa}$ and a Poisson's ratio of $\nu = 0.5$ for PDMS rubber, while
PMMA is treated as rigid. The vertical dashed lines indicate the average Hertzian line-contact pressures (nominal contact pressures) at the tested normal forces.

For Hertzian line contact, the maximal pressure is a factor of $4/\pi \approx 1.27$ larger than the average
pressure. Hence, Fig. \ref{1pressure.2AreaSoothRough1.eps} shows that if adhesion is negligible for the two sandblasted surfaces (blue and red lines), we expect the contact area to be proportional to pressure and thus to the normal force acting on the cylinder for
all studied normal forces. For the Smooth PMMA surface (green line), some nonlinearity in the dependence of contact area on normal force is expected.

\begin{figure}
\includegraphics[width=0.47\textwidth,angle=0.0]{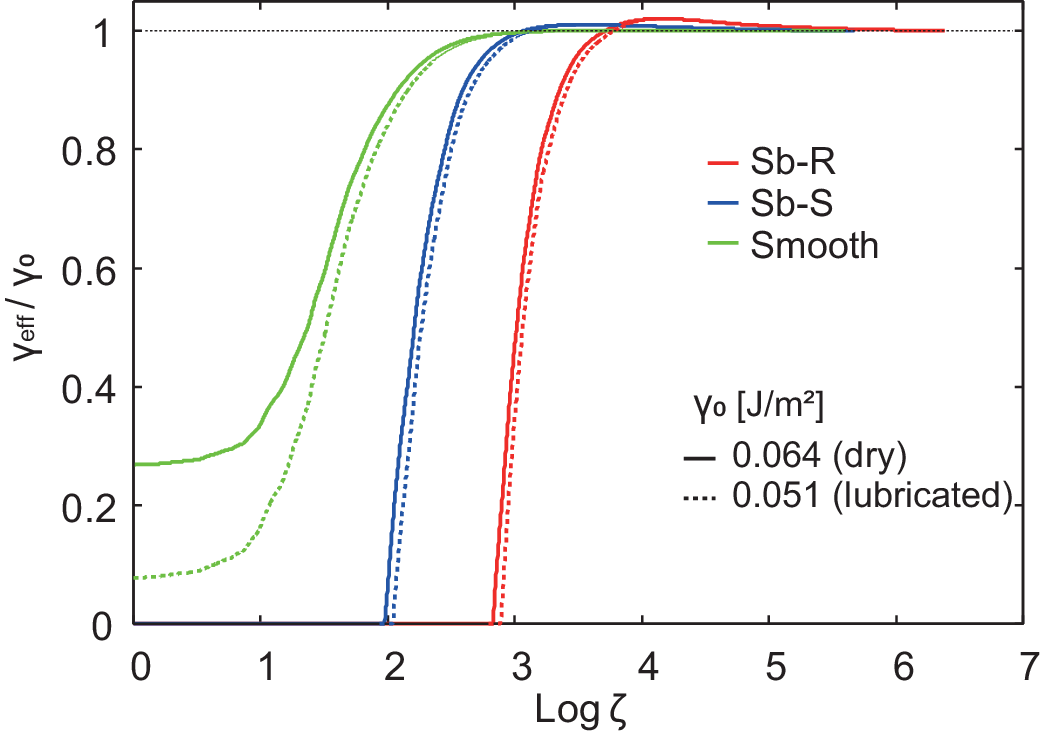}
\caption{\label{1LogZ.2Gammaeff.all.eps}
The effective binding energy per unit surface area, $\gamma_{\rm eff}$, as a function of magnification, normalized by $\gamma_0$, the binding energy for perfectly smooth surfaces. We used $\gamma_0 = 64$ and $51 \ {\rm mJ/m^2}$
for the dry contact and for the contact in glycerol, respectively.
}
\end{figure}

\begin{figure}
\includegraphics[width=0.47\textwidth,angle=0.0]{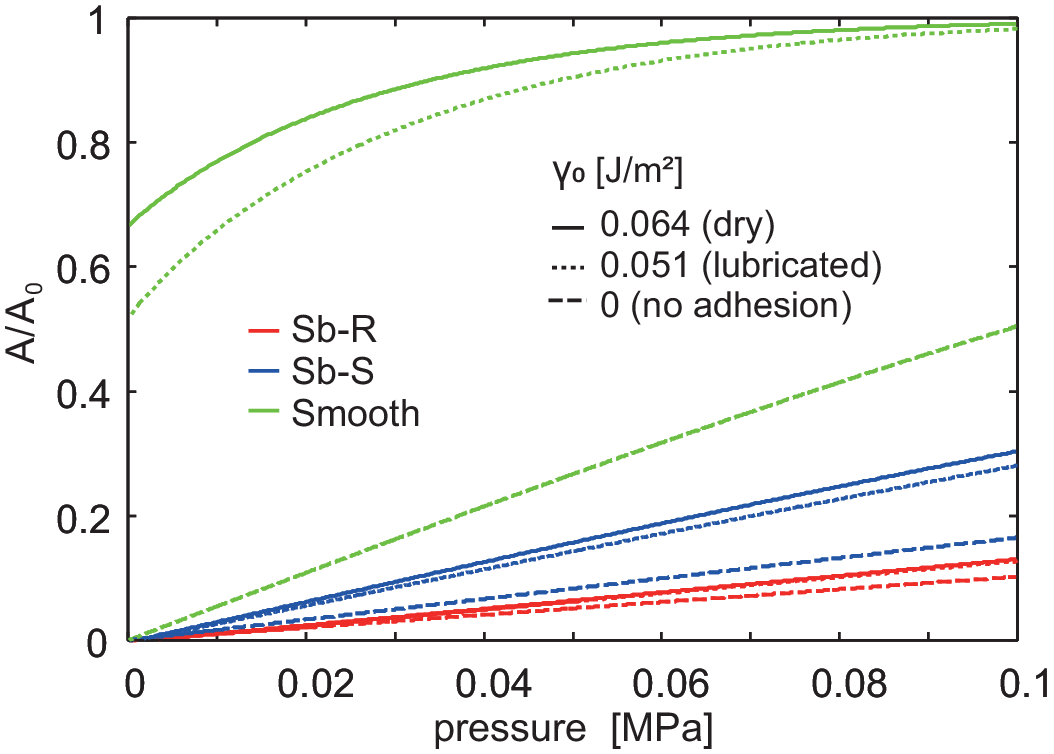}
\caption{\label{1PressureAreaAllNew.eps}
The relative contact area $A/A_0$ as a function of pressure for Smooth (green), Sb-S (blue), and Sb-R (red). The dashed lines represent the non-adhesive case, while the solid and dotted lines include adhesion in glycerol and in the dry state, respectively. Adhesion increases the contact area for all surfaces, with the strongest effect for Smooth.
}
\end{figure}

\vskip 0.3cm
\section{Adhesion enhances contact area}

The results for the contact area presented in Fig. \ref{1pressure.2AreaSoothRough1.eps} assume negligible adhesion.
For the Sb-S and Sb-R surfaces, when brought into contact with the dry PDMS surface,
we do not observe any adherence during pull-off (the pull-off force vanishes). 
However, for the Smooth surface we obtain a nonzero pull-off force.
These tests were performed manually rather than measured quantitatively, but the results agree
qualitatively with the adhesion calculations presented below.

We calculated the effective work of adhesion $\gamma_{\rm eff} (q)$ using the theory developed in Ref. \cite{PerssonAdh1}:
\begin{equation}
\begin{aligned}
\gamma_{\rm eff}(q) P(q) =\ & \gamma_0 \Big[ P(q_1) \int_0^\infty dx \, (1+\xi^2 x)^{1/2} e^{-x} \\
& - { 2\pi \over \delta} \int_q^{q_1} dq \, q^2 P(q) S(q) C(q) \Big]
\end{aligned}
\tag{11}
\end{equation}
where $\delta = 4 \gamma_0 / E^*$ is the adhesion length, and
$$\xi^2 = 2 \pi \int_{q_0}^{q_1} dq \, q^3 C(q)\eqno(12)$$
is the mean square surface slope. The function $P(q)=A(q)/A_0$ is the relative contact area when the interface is observed
at the magnification $\zeta = q/q_0$. For the non-adhesive case, $P(q)$ is given by (6) and (7), where in (7)
the upper limit of the $q$ integration is $q$ instead of $q_1$. When adhesion is non-negligible, $P(q)$ is determined by an integral
equation as described in Ref. \cite{PerssonAdh1}. The function $S(q)= \kappa+(1-\kappa)P^2(q)$, with $\kappa \approx 0.4$,
is an elastic energy reduction factor.

It is convenient to consider $\gamma_{\rm eff}(q)$ as a function of the magnification $\zeta = q/q_0$,
and write it as $\gamma_{\rm eff}(\zeta)$ for simplicity. If the effective work of adhesion at the macroscale satisfies $\gamma_{\rm eff}(1) > 0$,
then the pull-off force $F_{\rm pull} > 0$ and the real contact area $A > 0$ even when the applied pressure $p$
vanishes \cite{PerssonAdh1,PerssonAdh2}. However, if $\gamma_{\rm eff}(1) = 0$, then $F_{\rm pull} = 0$, and $A(p) \rightarrow 0$ as $p \rightarrow 0$.
Thus, for $p > 0$, the contribution of adhesion accumulates through the magnification-dependent contact area and increases the real contact area, and the sliding friction force {\it is always} enhanced by adhesion even when the adhesion does not manifest itself as a macroscopic pull-off force \cite{PerssonAdh3}.

In order to quantitatively study how adhesion depends on surface roughness, we need the interfacial binding energy per unit surface area
for smooth surfaces, which we denote by $\gamma_0$. This is also the work of adhesion required to adiabatically (infinitely slowly) separate the
surfaces. We are not aware of any measurements of this quantity for the dry and lubricated (with glycerol) PMMA-PDMS interface, and therefore estimate it for the dry contact using
$$\gamma_{\rm dry} = \gamma_{12} \approx 2 \surd (\gamma_1 \gamma_2)\eqno(13)$$
where $\gamma_1$ and $\gamma_2$ are the surface energies of the
PDMS and PMMA surfaces in the dry state.

For the lubricated interface, we consider the change in free energy as the surfaces are separated in glycerol.
The energy needed to break the bond between the surfaces is $\gamma_{\rm dry}$, but after separation the surfaces bind to glycerol.
Thus, the interfacial energy of a separated surface is not the solid-vapor energy $\gamma_{\rm sv}$ but the solid-liquid energy $\gamma_{\rm sl}$.
To account for this, we must subtract from $\gamma_{\rm dry}$ the difference
$\gamma_{\rm sv}-\gamma_{\rm sl} = \gamma {\rm cos}\theta$
for each surface, where we have used Young's relation. Thus, we obtain the standard result
$$\gamma_{\rm wet} = \gamma_{\rm dry}-\gamma ({\rm cos} \theta_1+{\rm cos} \theta_2)\eqno(14)$$
where $\theta_1$ and $\theta_2$ are the (adiabatic or thermal-equilibrium) contact angles of glycerol on the PDMS and PMMA surfaces,
respectively, and $\gamma$ is the surface tension of glycerol.

Here we use the glycerol surface tension $\gamma = 64 \ {\rm mJ/m^2}$, the surface energy of
PDMS, $\gamma_1 = 22 \ {\rm mJ/m^2}$, and the surface energy of PMMA, $\gamma_2 = 46 \ {\rm mJ/m^2}$.
For the contact angles of glycerol on PDMS and PMMA,
we use $\theta_1 = 104^\circ$ and $\theta_2 = 64^\circ$, respectively.

We note that the literature values for all these quantities vary. For example,
for PDMS we found surface energies in the range from $19.4$ to $24.3 \ {\rm mJ/m^2}$ \cite{SurfaceEnergyPDMS-Yasuri},
and for PMMA we found $33 \ {\rm mJ/m^2}$ \cite{PMMA-Abdel}, $40 \ {\rm mJ/m^2}$ \cite{PMMA-Cabanillas},
and $46 \ {\rm mJ/m^2}$ in Ref. \cite{web}. For the contact angle of glycerol on PDMS, we found $104^\circ$ \cite{PDMS-Zhang}
and $94-104^\circ$ \cite{Jagota}, while for glycerol on PMMA we found $\theta_2 = 68^\circ$ \cite{PMMA-Abdel},
$58-66.8^\circ$ \cite{Contactangles-Yasuri}, and $69^\circ$ \cite{PMMA-Cabanillas}.

Using the chosen values gives a work of adhesion for dry surfaces of
$\gamma_{\rm dry} \approx 2 \surd (\gamma_1 \gamma_2) \approx 64 \ {\rm mJ/m^2}$,
and a work of adhesion in glycerol of
$\gamma_{\rm wet} = \gamma_{\rm dry}-\gamma [{\rm cos} \theta_1+ {\rm cos} \theta_2 ] \approx 51 \ {\rm mJ/m^2}$.
However, due to the uncertainty in $\gamma_1$, $\gamma_2$, $\theta_1$, and $\theta_2$, and due to the approximate nature of the relation $\gamma_{\rm dry} \approx 2 \surd (\gamma_1 \gamma_2)$,
the values of $\gamma_{\rm dry}$ and $\gamma_{\rm wet}$ are only approximate.

Fig. \ref{1LogZ.2Gammaeff.all.eps} shows the dependence of the normalized effective binding energy per unit surface area,
$\gamma_{\rm eff}/\gamma_0$, on magnification for the Smooth, Sb-S and Sb-R surfaces, calculated using the theory developed in Ref. \cite{PerssonAdh1,PerssonAdh2}. The solid lines are for the dry contact with $\gamma_0 = \gamma_{\rm dry} = 64 \ {\rm mJ/m^2}$,
and the dotted lines are for the contact in glycerol with $\gamma_0 = \gamma_{\rm wet} = 51 \ {\rm mJ/m^2}$.
Note that for Sb-S and Sb-R, the macroscopic work of adhesion vanishes, $\gamma_{\rm eff}(1)=0$.
This is not the case for the Smooth surface, so for this surface
the pull-off force is nonzero, as indeed observed.

Fig. \ref{1PressureAreaAllNew.eps} shows the dependence of the contact area on the applied pressure.
The dashed lines represent the non-adhesive case, while the solid and dotted lines include adhesion in the dry state and in glycerol, respectively.
For all three surfaces, adhesion increases the contact area at finite pressure. This increase is strongest for the Smooth surface, moderate for Sb-S, and weakest for Sb-R.

As expected, the contact area remains finite for the Smooth surface as $p \rightarrow 0$ both in the dry state
and in glycerol. In contrast, for Sb-S and Sb-R, the contact area at $p=0$ vanishes in both the dry state and in glycerol.
Nevertheless, for all surfaces, adhesion enhances the contact area at finite pressures \cite{P1}.

This trend is confirmed by the contact areas at the lowest nominal contact pressure tested, $\sigma_0 = 0.042 \ {\rm MPa}$. For the Smooth surface, the contact areas without adhesion, with adhesion in glycerol, and in the dry state are
$A/A_0 = 0.224$, $0.876$, and $0.924$, respectively. For Sb-S, the corresponding values are $A/A_0 = 0.069$, $0.119$, and $0.131$,
and for Sb-R, $A/A_0 = 0.043$, $0.052$, and $0.053$.

The same figure also shows that for Sb-R the contact area in glycerol (solid red line) is nearly
the same as in the dry state (dotted red line). For Sb-S (blue lines), the contact area in the dry state is about
$8 \%$ larger than in glycerol. For all surfaces, the contact area, including adhesion is considerably larger than that in the non-adhesive case
(dashed lines).

We summarize the most important results related to adhesion:

(a) If $\gamma_{\rm eff}(1)>0$, adhesion will increase the size of the nominal contact region. In the present case of a cylinder-flat
line contact, the half-width $a$ of the contact is determined by \cite{chaud}:
$${F\over L} = {\pi E^*a^2 \over 4R}-(2E^*\pi a w )^{1/2}\eqno(15)$$
where $w=\gamma_{\rm eff}(1)$ is the macroscopic work of adhesion. For Sb-S and Sb-R, $w=0$, and the half-width $a$ is given by
Hertz theory:
$$a = \left({4RF\over \pi E^*L}\right)^{1/2} \eqno(16)$$
For the Smooth surface, $w>0$, but the effect of adhesion on the nominal contact width is small even for the smallest load
$F/L = 310 \ {\rm N/m}$, where $a$ increases only from $37 \ {\rm mm}$ for $w=0$ to $40 \ {\rm mm}$ for $w=0.064 \ {\rm J/m^2}$.

(b) Adhesion {\it always} increases the relative contact area $A/A_0$ {\it inside} the nominal contact area. This is the case even when $\gamma_{\rm eff}(1)=0$.
Hence, adhesion will {\it always} increase the real contact area.

(c) If $\gamma_{\rm eff}(1)=0$, the pull-off force vanishes, while it is finite when $\gamma_{\rm eff}(1)>0$.

(d) If $\gamma_{\rm eff}(1)=0$, there can be no macroscopic elastic instabilities, i.e., no instabilities resulting from adhesion on the length scale of the linear size of the system, and hence no Schallamach waves \cite{Schallwaves}. However, adhesion still acts at shorter length scales, so it is still possible that elastic instabilities occur at shorter length scales. One such process is the origin of the adhesive contribution to sliding friction, involving nanosized patches of the rubber binding to the counter surface and performing stick-slip-stick-type motion during sliding. This is the Schallamach mechanism \cite{Schall} for the origin of the adhesive contribution to rubber friction, which was studied using a more realistic model by Persson and Volokitin \cite{PVSchall}.


\begin{figure}
\includegraphics[width=0.47\textwidth,angle=0.0]{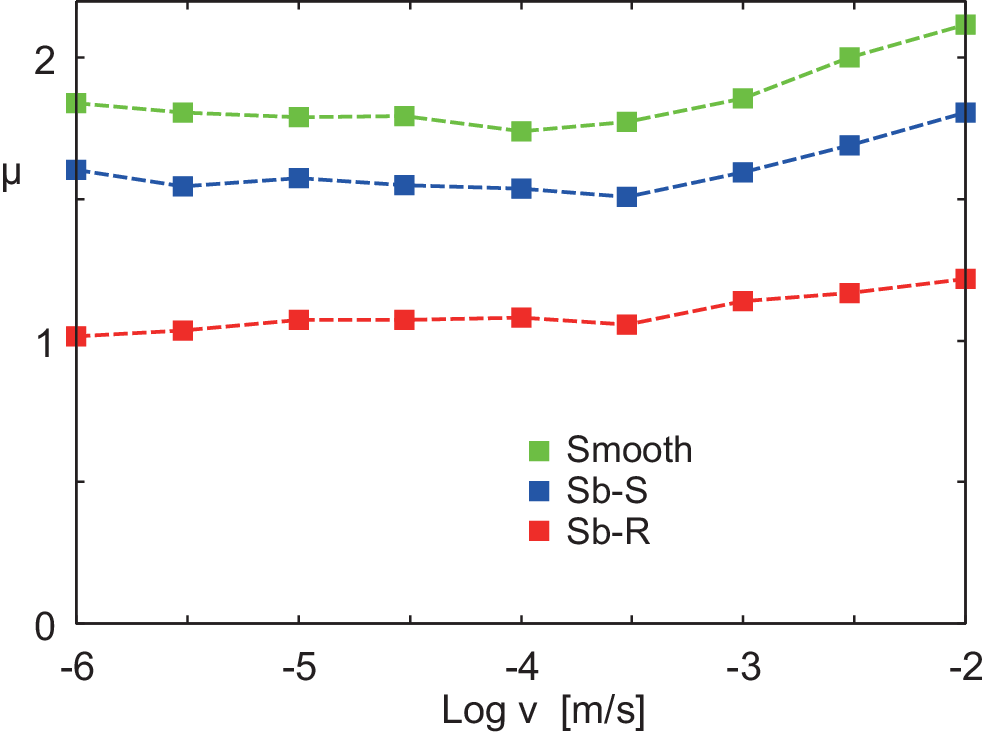}
\caption{\label{1logv.2mu.experiment.dy.all.eps}
The measured dry friction coefficient as a function of sliding speed
for the Smooth, Sb-S, and Sb-R surfaces sliding on PDMS.
In the experiments, the nominal contact pressure was $\sigma_0=0.042 \ {\rm MPa}$.
}
\end{figure}

\begin{figure}
\includegraphics[width=0.47\textwidth,angle=0.0]{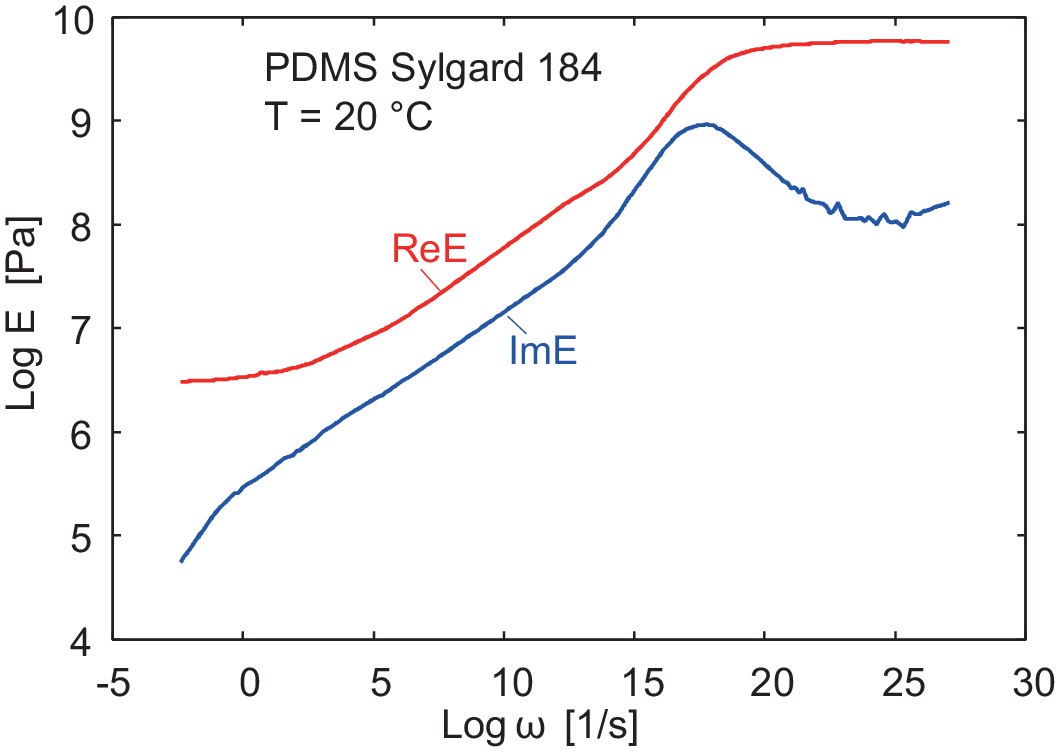}
\caption{\label{1logOmega.2logE.eps}
The real and imaginary parts of the viscoelastic modulus of the PDMS rubber
as a function of angular frequency (log-log scale) at the temperature
$T=20^\circ {\rm C}$.
}
\end{figure}

\begin{figure}
\includegraphics[width=0.47\textwidth,angle=0.0]{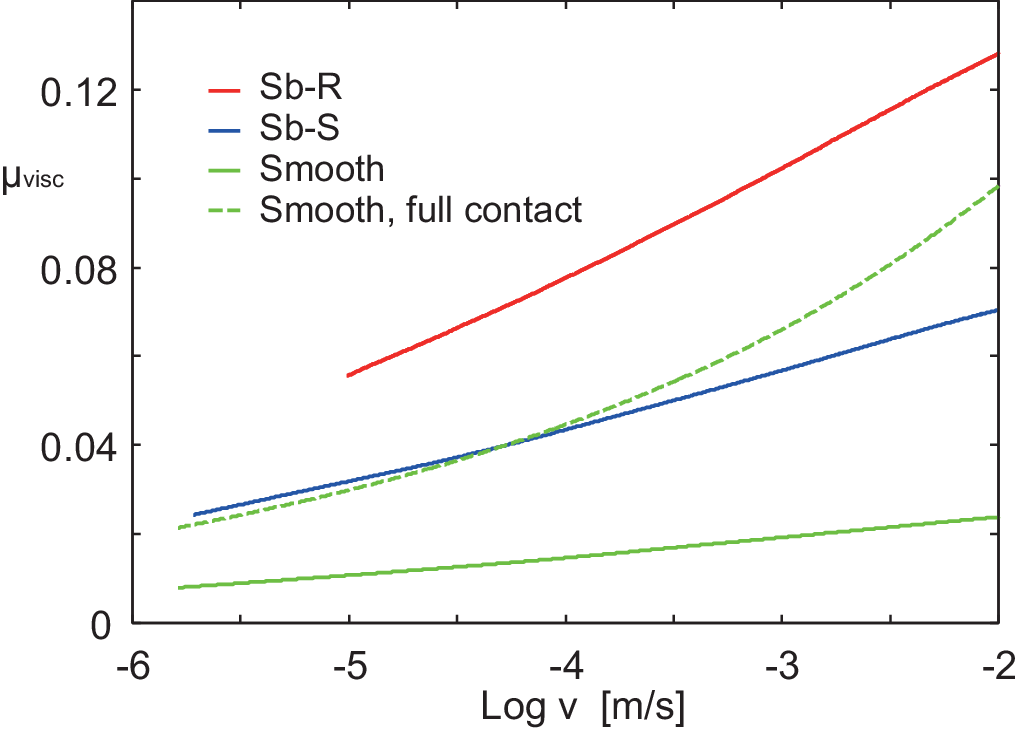}
\caption{\label{1logv.2mu.surface1and2.eps}
The calculated viscoelastic contribution to the
sliding friction coefficient for PDMS rubber sliding on the Smooth (green), Sb-S (blue), and Sb-R (red) surfaces.
The solid lines are the results obtained without adhesion.
The dashed green line is the friction coefficient for the Smooth surface
assuming complete contact between the surfaces due to adhesion.
At the temperature $T=20^\circ {\rm C}$ and the nominal contact pressure $\sigma_0 = 0.042 \ {\rm MPa}$.
}
\end{figure}

\begin{figure}
\includegraphics[width=0.47\textwidth,angle=0.0]{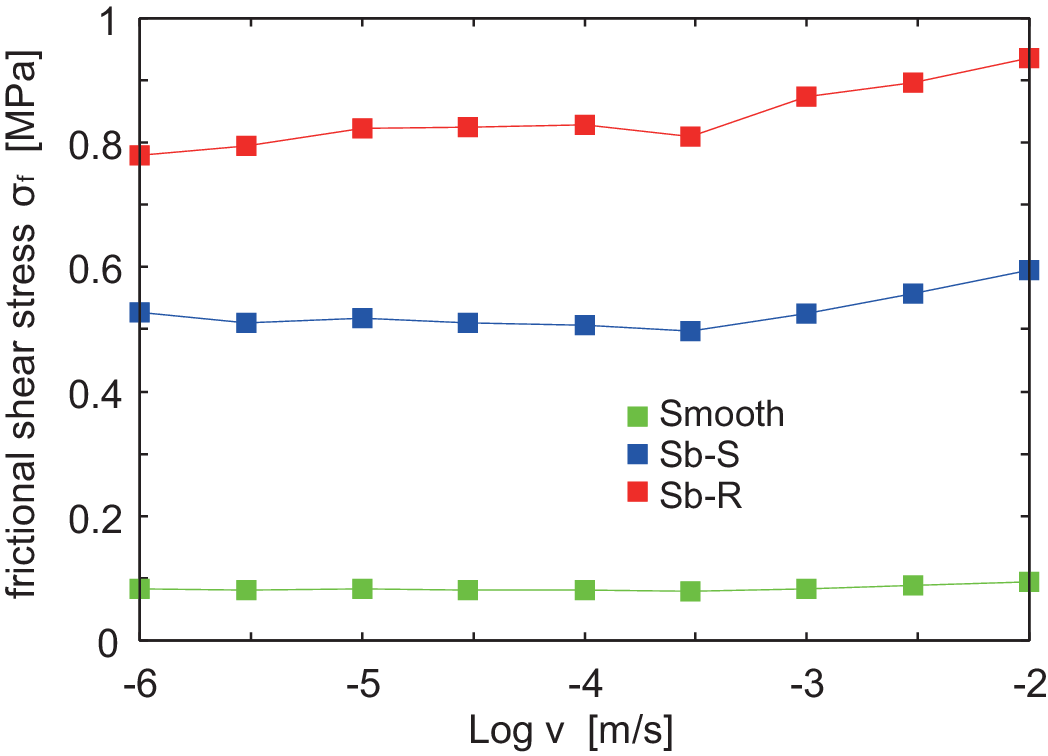}
\caption{\label{1logv.2shearstress.from.mu.and.area.from.adhesion.eps}
The effective frictional shear stress $\sigma_{\rm f}$ in the dry state, calculated by assuming that the friction force satisfies
$\sigma_{\rm f} A = \mu \sigma_0 A_0$, where $\mu$ is given in Fig. \ref{1logv.2mu.experiment.dy.all.eps},
and using the dry contact areas at $\sigma_0 =0.042 \ {\rm MPa}$ from Fig. \ref{1PressureAreaAllNew.eps}:
$A/A_0 \approx 0.924$ for the Smooth surface, $\approx 0.131$
for Sb-S, and $\approx 0.053$ for Sb-R.
}
\end{figure}

\vskip 0.3cm
\section{Dry friction and interfacial shear stress}

The analysis above shows that the real contact area can be significantly increased by adhesion, especially for the Smooth surface. This has important consequences for the interpretation of friction measurements. If adhesion is not included explicitly in the friction model, then the frictional shear stress deduced from the measured friction force should be regarded as an effective shear stress, which is larger than the actual shear stress acting in the real contact area. Assuming that the adhesion-induced enhancement of the contact area is reasonably described by the static calculations presented above, one may estimate the actual shear stress by correcting for the increase in contact area.

With this in mind, we first analyze dry friction in order to identify the origin and relative magnitude of the interfacial shear stress before turning to the lubricated Stribeck curves.

The green, blue, and red squares in Fig. \ref{1logv.2mu.experiment.dy.all.eps}
show the measured dry friction coefficient as a function of sliding speed for the Smooth, Sb-S, and Sb-R surfaces, respectively.
The results are for the normal load per unit length $F_{\rm N}/L = 310 \ {\rm N/m}$,
and the nominal contact pressure $\sigma_0 = 0.042\ {\rm MPa}$.

The viscoelastic contribution is calculated using the Persson rubber friction theory. We include the surface roughness
on all length scales from $\sim 1 \ {\rm cm}$ to $\sim 1 \ {\rm nm}$ using the surface roughness power spectra shown in Fig.
\ref{cut1.and.not.logq.2logC.all.full.eps}. In the calculations we use the measured viscoelastic modulus shown in Fig. \ref{1logOmega.2logE.eps}. We note that, because of
crystallization of the PDMS during cooling, it is nontrivial to determine the viscoelastic modulus over the full frequency range using
the temperature-frequency shifting procedure, and the results shown in Fig. \ref{1logOmega.2logE.eps} were obtained by
interpolating between the rubbery and glassy frequency regions as discussed in Ref. \cite{PerssonAdh3}.

Fig. \ref{1logOmega.2logE.eps} shows the real and imaginary parts of the viscoelastic modulus of the PDMS rubber
as a function of angular frequency (log-log scale) at the temperature
$T=20^\circ {\rm C}$. Using this modulus and the surface roughness power spectra shown in Fig.
\ref{cut1.and.not.logq.2logC.all.full.eps}, the solid lines in Fig. \ref{1logv.2mu.surface1and2.eps}
show the calculated viscoelastic contribution to the sliding friction coefficient for PDMS rubber sliding on the
Smooth surface and on the sandblasted PMMA surfaces Sb-S and Sb-R.
The calculations do not include the increase in $A/A_0$ due to adhesion. 

As a simple estimate, we assume that the viscoelastic contribution is proportional to the contact area obtained in the static adhesion calculations.
A more rigorous treatment would require including the adhesion-induced change in the relative contact area at each magnification, since different length scales contribute differently to the viscoelastic dissipation.
The present estimate should therefore be regarded as an upper-bound estimate: scaling the viscoelastic friction with the adhesion-induced increase in the real contact area tends to overestimate the influence of adhesion, since the enhancement of the apparent contact area decreases as the magnification is reduced.
With this linear scaling assumption, adhesion increases the viscoelastic friction by factors of
$\approx 0.1313/0.0692 \approx 1.89$ and $0.0528/0.0428 \approx 1.23$ for Sb-S and Sb-R, respectively. This would result in similar
viscoelastic friction coefficients for both surfaces, $\mu_{\rm visc} \sim 0.1$, which is much smaller than observed.
For the Smooth surface, the adhesion is so strong that it nearly pulls the surfaces into complete contact,
and the dashed green line in Fig. \ref{1logv.2mu.surface1and2.eps} is obtained by assuming complete contact,
$A/A_0 = 1$. In this case as well, the viscoelastic contribution is at most $\sim 0.1$.
Hence, we conclude that the friction for PDMS sliding on all surfaces is mainly due to adhesive molecular interactions, i.e.,
the formation and breaking of adhesive bonds between the PMMA and the silicone rubber
at the sliding interface. The weak dependence of friction on sliding speed
shown in Fig. \ref{1logv.2mu.surface1and2.eps} indicates that thermal activation is not very important for the bond-breaking (or bond-formation)
process, which is expected if the energies involved are much larger than the thermal energy.

From the measured friction coefficients in Fig. \ref{1logv.2mu.experiment.dy.all.eps}, using
$\sigma_{\rm f} A = \mu \sigma_0 A_0$, we can estimate an effective frictional shear stress associated with the
real contact area. Here $A/A_0$ is the relative contact area
including adhesion (see Fig. \ref{1PressureAreaAllNew.eps}). The quantity $\sigma_{\rm f}$ should be viewed as an effective interfacial shear stress which may include several dissipative contributions, such as molecular shear in the real contact area, dissipation near the edges of contact patches, and local stick-slip processes.
For $\sigma_0 = 0.042 \ {\rm MPa}$, we have
$A/A_0 = 0.924$, $0.131$, and $0.053$ for the Smooth surface, Sb-S, and Sb-R, respectively.
Using this, we obtain the $\sigma_{\rm f}(v)$ results shown in Fig. \ref{1logv.2shearstress.from.mu.and.area.from.adhesion.eps}. This figure shows that the effective frictional shear stress for Sb-R is a factor of $\sim 1.5$ larger than for Sb-S, and both are much larger than
that for the Smooth surface. In particular, the much smaller $\sigma_{\rm f}$ for the Smooth surface suggests that the sliding mode differs from that for Sb-S and Sb-R. Although the present experimental setup does not allow direct observation of the sliding interface, the friction traces provide indirect evidence for this difference.

\begin{figure}
\includegraphics[width=0.45\textwidth,angle=0.0]{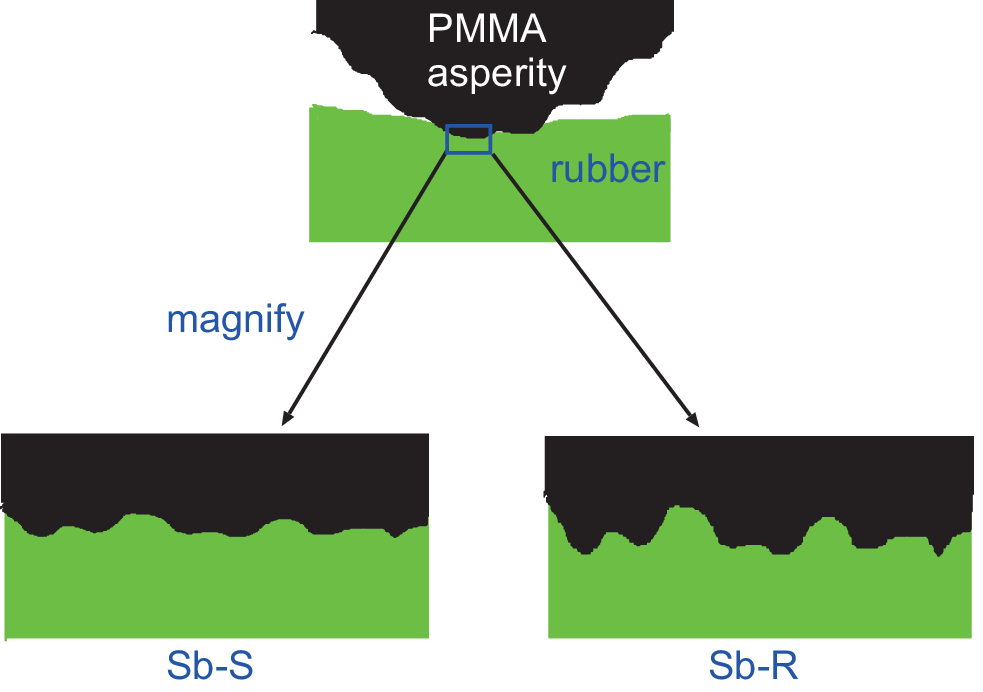}
\caption{\label{binding.eps}
Schematic illustration showing that the frictional shear stress in the real contact area may be larger for the
surface with greater short-wavelength roughness, because stronger bonds
can form between nanosized rubber patches and the counter surface.
}
\end{figure}

The larger $\sigma_{\rm f}$ for Sb-R than for Sb-S can be understood using the generalized Schallamach picture of the adhesive contribution to sliding friction \cite{Schall,PVSchall}. In this picture, friction results from nanosized patches of rubber at the interface performing stick-slip (binding-stretching-detachment-binding) motion. The larger short-wavelength roughness of Sb-R can make the real contact area more fragmented, with a larger number of smaller contact spots and contact edges, as indicated schematically in Fig. \ref{binding.eps}. This contact morphology may enhance dissipation associated with local stick-slip and with the edges of adhesive contact patches. It may also increase local pinning barriers in the sliding direction. These effects provide a possible explanation for the larger effective interfacial shear stress observed for Sb-R than for Sb-S.


This interpretation is consistent with earlier observations by Chateauminois and Fretigny \cite{ShearStress}
(see also, \cite{Jacobs,Chaudhury}). They measured the shear stress distribution at a frictional interface between a flat PDMS substrate and a glass lens. For a smooth glass lens (rms roughness less than 2 nm), they found that the frictional shear stress in the nominal contact region is nearly constant, indicating that it does not depend on the local contact pressure when the contact is essentially complete and the pressure is not too high. For the same lens after sandblasting (rms roughness $\sim 1 \ {\rm \mu m}$ over a $0.1 \times 0.1 \ {\rm mm}$ region), the shear stress in the central part of the Hertz-like contact remained independent of load, but its magnitude increased by about a factor of 2 compared to the smooth surface. In an outer annular region, however, the shear stress depended on the local contact pressure, indicating that the pressure there was not high enough to squeeze the rubber into complete contact with the glass lens and that only partial contact occurred. These observations support the idea that short-wavelength roughness can increase the local frictional shear stress.

\begin{figure}
\includegraphics[width=0.47\textwidth,angle=0.0]{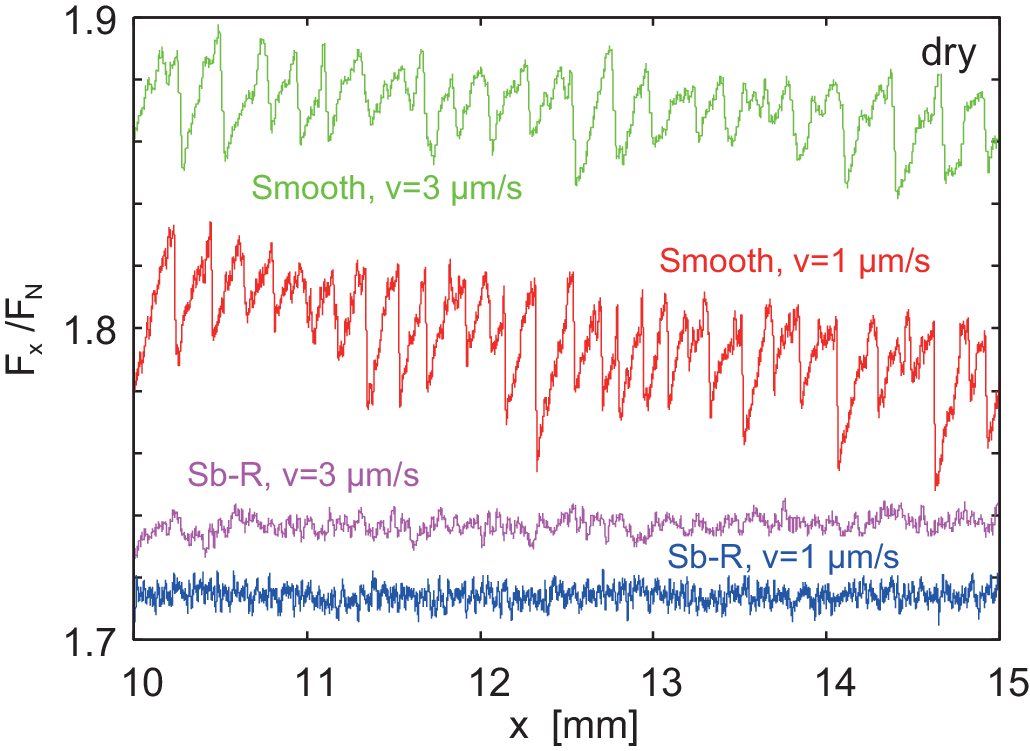}
\caption{\label{DRY.eps}
The ratio $\mu = F_x/F_{\rm N}$ as a function of sliding distance for the Smooth
(upper two curves) and Sb-R (lower two curves) cylinder on the dry
PDMS surface. The sliding speeds are $1$ and $3 \ {\rm \mu m/s}$, and the
normal load is $F_{\rm N}/L = 310 \ {\rm N/m}$. The green, blue, and purple curves are shifted vertically for clarity.
Note the stick-slip oscillations in the friction force
for the Smooth surface.
}
\end{figure}

Fig. \ref{DRY.eps} shows the ratio $\mu = F_x/F_{\rm N}$ as a function of sliding distance for the Smooth surface (upper two curves) and Sb-R (lower two curves) on the dry PDMS surface. The sliding speeds $v$ are $1$ and $3 \ {\rm \mu m/s}$, and the
normal load is $F_{\rm N}/L = 310 \ {\rm N/m}$. Note the stick-slip oscillations in the friction force
for the Smooth surface. The friction curves for the Smooth surface are very similar in both velocities. 
These observations are consistent with the occurrence of an instability-mediated sliding mode for the Smooth surface. 
The physical origin of these differences will be discussed later.


\begin{figure}
\includegraphics[width=0.47\textwidth,angle=0.0]{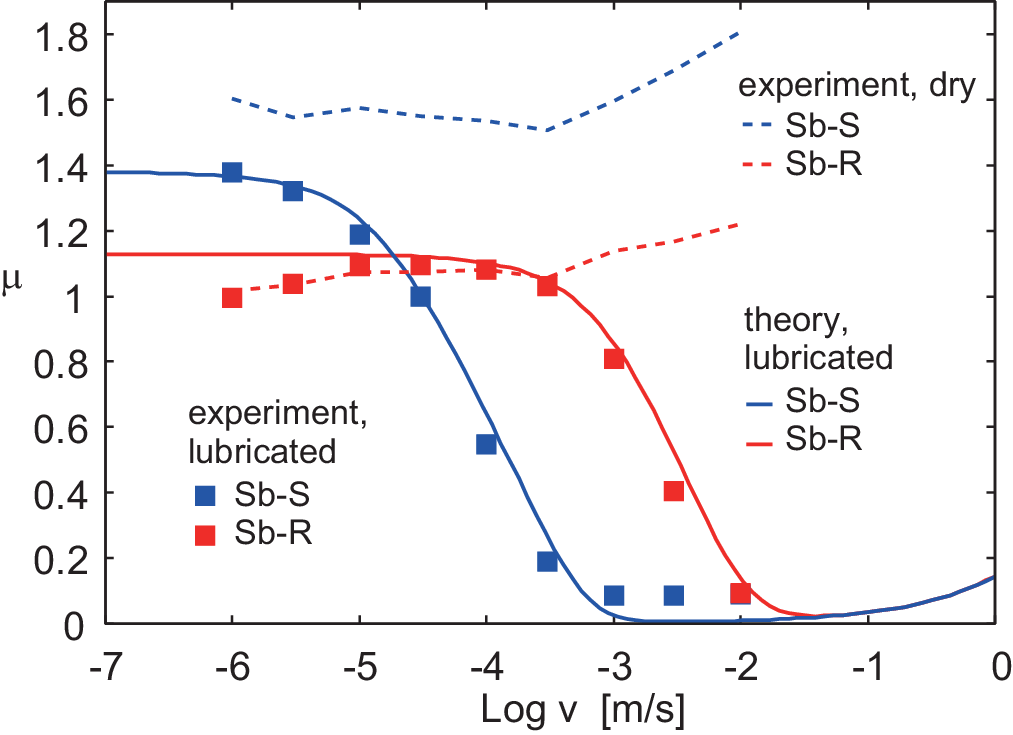}
\caption{\label{1logv.2mu.smooth0.0.84.rough0.1.08.exp.eta.eps}
The friction coefficient as a function of the logarithm of sliding speed for Sb-S and Sb-R. The square symbols and solid lines are the measured and calculated friction coefficients for glycerol-lubricated contacts, respectively, and the dashed lines are the corresponding dry-friction results (from Fig. \ref{1logv.2mu.experiment.dy.all.eps}). The calculations use a fluid viscosity $\eta = 1.4 \ {\rm Pa\,s}$, a cylinder radius $R = 7 \ {\rm cm}$, a normal force per unit length $F_{\rm N}/L = 310 \ {\rm N/m}$, and fitted shear stresses $\sigma_{\rm f} = 0.84$ and $1.08 \ {\rm MPa}$.
}
\end{figure}

\begin{figure}
\includegraphics[width=0.47\textwidth,angle=0.0]{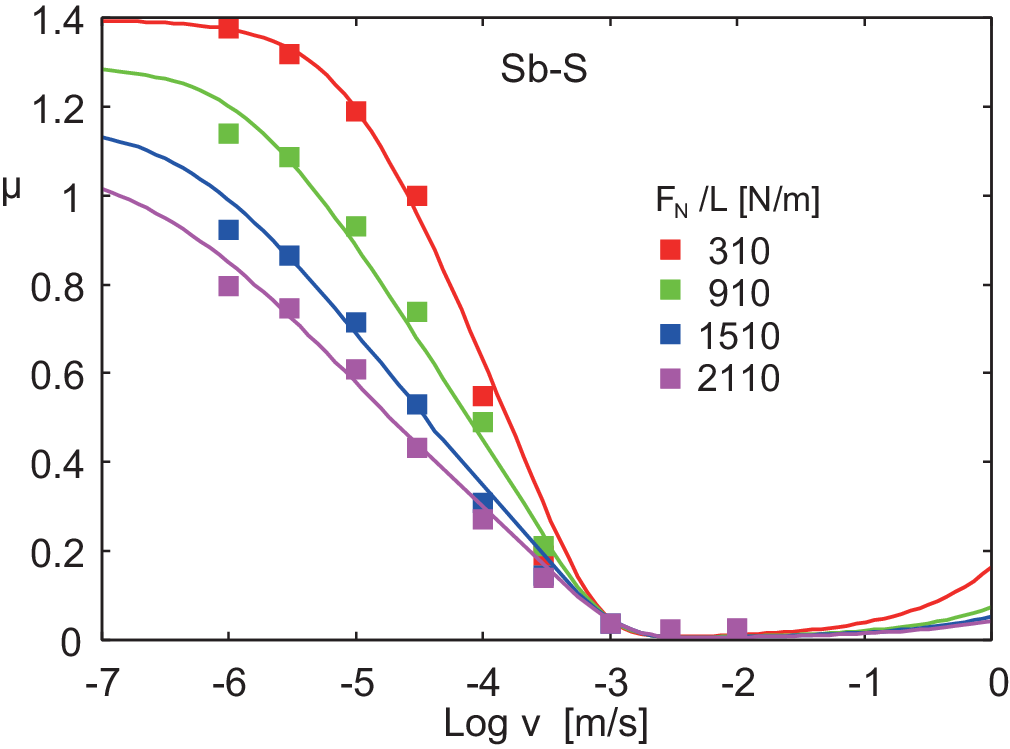}
\caption{\label{1logv.2mu.smooth.all.pressures.a0.84.b0.78.c0.7.d0.65.eps}
The friction coefficient as a function of the logarithm of sliding speed for Sb-S at different normal loads. The square symbols and lines are the measured and calculated results, respectively, for $F_{\rm N}/L = 310$, $910$, $1510$, and $2110 \ {\rm N/m}$. The corresponding fitted velocity-independent shear stresses for increasing load are $\sigma_{\rm f} = 0.84$, $0.78$, $0.7$, and $0.65 \ {\rm MPa}$.
}
\end{figure}

\begin{figure}
\includegraphics[width=0.47\textwidth,angle=0.0]{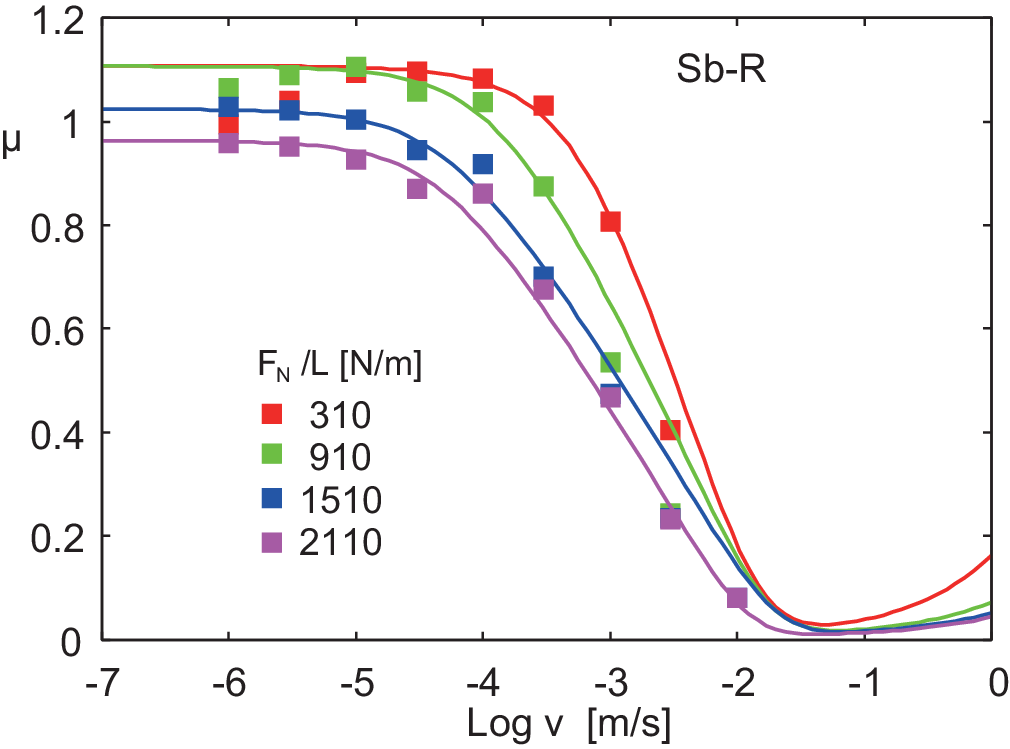}
\caption{\label{1logv.2mu.rough.all.loads.a1.08.b1.08.c1.0.d0.94.eps}
The friction coefficient as a function of the logarithm of sliding speed for Sb-R at different normal loads. The square symbols and lines are the measured and calculated results, respectively, for $F_{\rm N}/L = 310$, $910$, $1510$, and $2110 \ {\rm N/m}$. The corresponding fitted velocity-independent shear stresses for increasing load are $\sigma_{\rm f} = 1.08$, $1.08$, $1.0$, and $0.98 \ {\rm MPa}$.
}
\end{figure}

\vskip 0.3cm
\section{Stribeck curves: experiment and theory}

The dry friction results above provide a baseline for the interfacial shear mechanism. We now turn to the lubricated case, where this interfacial contribution is coupled to fluid flow and load sharing between the fluid and the solid-solid contact.

Fig. \ref{1logv.2mu.smooth0.0.84.rough0.1.08.exp.eta.eps}
shows the friction coefficient as a function of the logarithm of sliding speed for the
two sandblasted PMMA cylinders. The fluid viscosity is $\eta = 1.4 \ {\rm Pa\,s}$,
and the normal force per unit length of the cylinder is $F_{\rm N}/L = 310 \ {\rm N/m}$. The blue and red squares are the measured
results for Sb-S and Sb-R, respectively, and the solid lines are the calculated friction coefficients obtained by assuming that a
velocity-independent frictional shear stress $\sigma_{\rm f}$ acts in the real contact area. For Sb-S and Sb-R we used fitted values of $\sigma_{\rm f} = 0.84$
and $1.08 \ {\rm MPa}$, respectively.

\begin{figure}
\includegraphics[width=0.47\textwidth,angle=0.0]{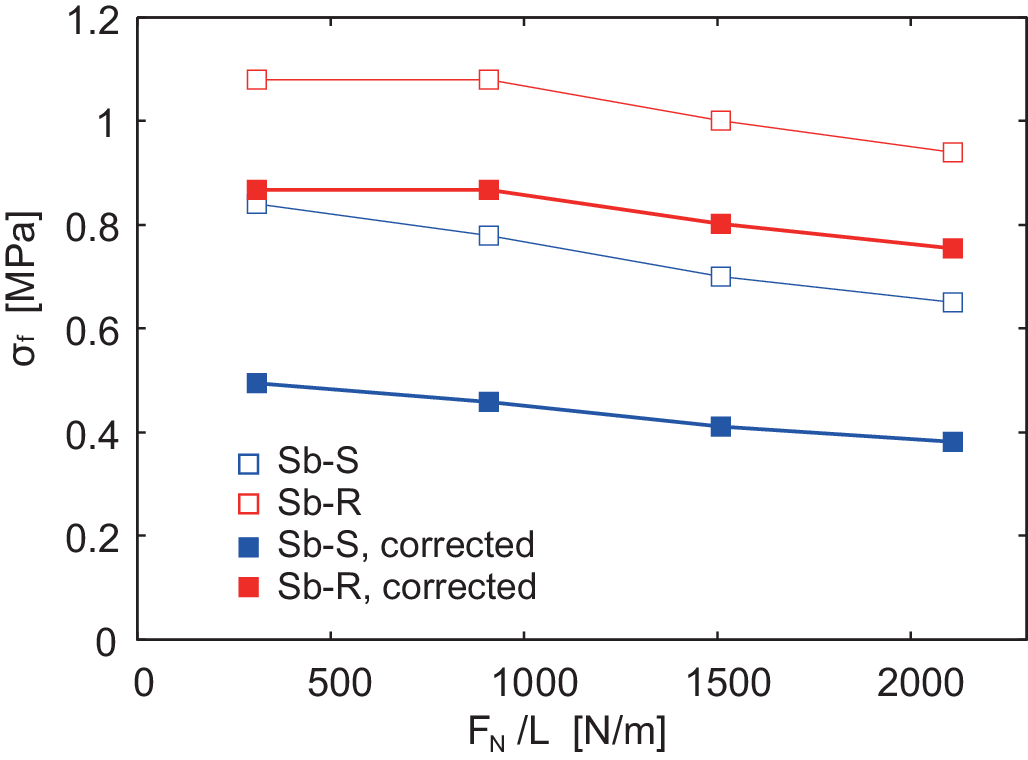}
\caption{\label{1Load.2sigmaf.both.eps}
The open squares are the frictional shear stress for Sb-S (blue squares) and Sb-R (red squares)
as a function of normal load, obtained by fitting the measured friction coefficients to the
calculated Stribeck curves in Fig. \ref{1logv.2mu.smooth.all.pressures.a0.84.b0.78.c0.7.d0.65.eps}
and \ref{1logv.2mu.rough.all.loads.a1.08.b1.08.c1.0.d0.94.eps}.
The filled square symbols are the ``corrected'' shear stresses, which take into account the
adhesion-induced increase in contact area. See the text for details.
}
\end{figure}

Figs. \ref{1logv.2mu.smooth.all.pressures.a0.84.b0.78.c0.7.d0.65.eps} and \ref{1logv.2mu.rough.all.loads.a1.08.b1.08.c1.0.d0.94.eps}
show the dependence of the Stribeck curve on the applied normal force.
The square symbols are the measured friction coefficients as a function of the logarithm of sliding speed
for the tested normal forces, and the lines are the corresponding calculated curves.
For Sb-S, the fitted values of $\sigma_{\rm f}$ for increasing load are $0.84$, $0.78$, $0.7$, and $0.65 \ {\rm MPa}$,
and for Sb-R they are $1.08$, $1.08$, $1.0$, and $0.98 \ {\rm MPa}$.
These fitted shear stresses are plotted in Fig. \ref{1Load.2sigmaf.both.eps} (open squares) as a function of normal load.

The fitted frictional shear stresses are systematically larger for Sb-R than for Sb-S. These values should be regarded as effective parameters within the present mixed-lubrication framework, since the calculations do not include the adhesion-induced increase in contact area and since the interfacial shear stress may include several dissipative contributions. The dry-friction results indicate that, for Sb-S and Sb-R, the effective interfacial shear stress may depend on sliding speed. In principle, these dry-friction values could be used in the boundary-lubrication regime after correcting for the adhesion-induced change in contact area. Here we use a velocity-independent $\sigma_{\rm f}$ as a minimal effective parameter, representing an average shear stress over the velocity range where solid-solid contact contributes to friction. Therefore, the fitted values are not expected to coincide quantitatively with the effective shear stresses inferred from the dry friction measurements. Nevertheless, the dry-friction results provide a useful baseline for the roughness dependence and order of magnitude of the interfacial shear contribution.

For the static contacts in glycerol, the contact area for Sb-R is nearly the same as in the dry state and larger by a factor of $0.127/0.102 \approx 1.25$ than without adhesion. For Sb-S, the contact area in glycerol is slightly smaller than for the dry contact and about $0.281/0.165 \approx 1.7$ larger than without adhesion. Scaling the shear stress given by the open square symbols in Fig. \ref{1Load.2sigmaf.both.eps}
by the factors $1/1.25$ for Sb-R and $1/1.7$ for Sb-S gives the filled square symbols in Fig. \ref{1Load.2sigmaf.both.eps}. The corrected ratio between the shear stresses for Sb-R and Sb-S varies slightly with load but remains between $1.75$ and $1.95$. This is somewhat larger than what was found for the dry contact, where it was $\sim 1.5$, but the difference may reflect uncertainties in the adhesion calculation.

\begin{figure}
\includegraphics[width=0.47\textwidth,angle=0.0]{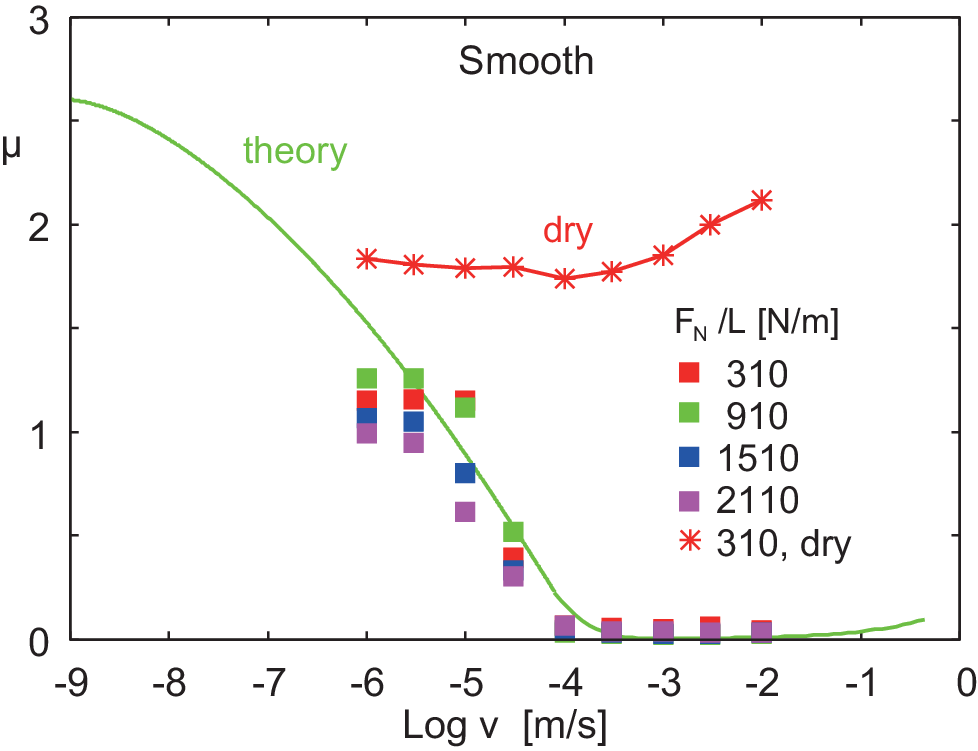}
\caption{\label{1logv.2mu.flat.everything.eps}
The square symbols are the measured friction coefficient as a function of the logarithm of sliding speed for the Smooth surface
for the loads $F_{\rm N}/L = 310$, $910$, $1510$, and $2110 \ {\rm N/m}$. The
green line is the calculated friction coefficient use $F_{\rm N}/L = 310 \ {\rm N/m}$, $\eta = 1.4 \ {\rm Pa\,s}$, $R = 7 \ {\rm cm}$, and assuming a velocity-independent frictional shear stress $\sigma_{\rm f} = 0.5 \ {\rm MPa}$. 
}
\end{figure}

We next consider the Smooth surface, the square symbols in Fig. \ref{1logv.2mu.flat.everything.eps} are the measured friction coefficients as a function of the logarithm of sliding speed.
The green line is the calculated friction coefficient assuming a velocity-independent frictional
shear stress $\sigma_{\rm f} = 0.5 \ {\rm MPa}$ and the load $F_{\rm N}/L = 310 \ {\rm N/m}$.
Since adhesion is not included in the analysis, the real contact area is larger, and the effective frictional shear stress is smaller.
Assuming that the contact area is enhanced by a factor of $0.876/0.224 \approx 3.91$, as predicted by the static adhesion calculation in Sec. 5,
the actual shear stress becomes $\sigma_{\rm f} = 0.127 \ {\rm MPa}$, which is similar to what we found for the dry contact.
The theory predicts that the boundary lubrication region is reached only at extremely low sliding speeds, below
$10^{-9} \ {\rm m/s}$, but the experiments show that the friction coefficient reaches a constant level (plateau) at much higher sliding speeds.

This discrepancy suggests a different sliding state for the Smooth surface at low sliding speeds, where direct solid-solid contact and adhesion are still important. As the sliding speed increases above $10 \ {\rm \mu m/s}$, forced wetting occurs and a fluid film enters the nominal contact region between the rubber and the cylinder. The transition results from the competition between liquid invasion induced by shear and spontaneous dewetting of the liquid between nonwettable surfaces \cite{dewet0,dewet1}.

\begin{figure}
\includegraphics[width=0.47\textwidth,angle=0.0]{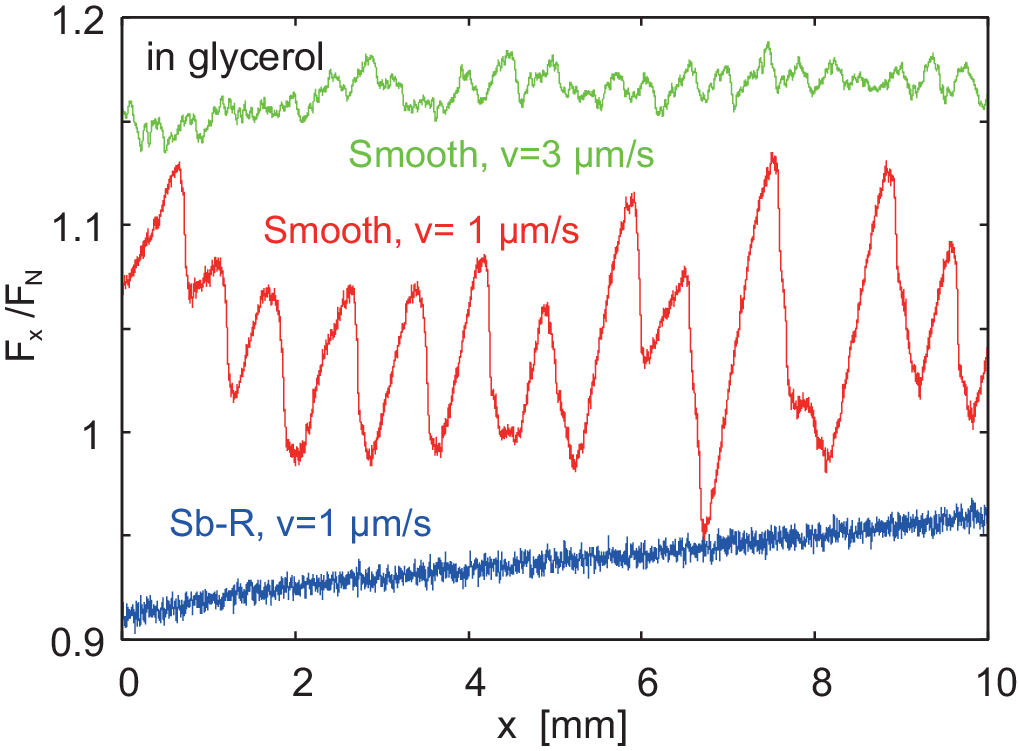}
\caption{\label{WET.eps}
The ratio $\mu = F_x/F_{\rm N}$ as a function of sliding distance for the Smooth surface (upper two curves) and Sb-R (lower curve) on the PDMS surface lubricated by glycerol. The sliding speeds are $v=1$ and $3 \ {\rm \mu m/s}$ for the Smooth surface and $v=1 \ {\rm \mu m/s}$ for Sb-R, and the
normal load is $F_{\rm N}/L = 310 \ {\rm N/m}$. Note the stick-slip oscillations in the friction force
for the Smooth surface. The green and blue curves are shifted vertically for clarity.
}
\end{figure}

Fig. \ref{WET.eps} shows the measured ratio $\mu = F_x/F_{\rm N}$ as a function of sliding distance for the Smooth
PMMA cylinder (upper two curves) and Sb-R (lower curve) on the
PDMS surface lubricated by glycerol. The sliding speed is $v=1 \ {\rm \mu m/s}$ for Sb-R and
$v=1$ and $3 \ {\rm \mu m/s}$ for the Smooth surface, at the
normal load of $F_{\rm N}/L = 310 \ {\rm N/m}$. Note the stick-slip oscillations in the friction force
for the Smooth surface. Compared to the dry contact, these oscillations have a longer wavelength and are more sinus-like than the
abrupt stick-slip pattern observed in the dry case. We attribute this to the influence
of glycerol on the formation and propagation of Schallamach waves. Note also that the stick-slip oscillations
for the Smooth surface are much larger at the sliding speed $1 \ {\rm \mu m/s}$
than at $v=3 \ {\rm \mu m/s}$ which we attribute to forced wetting.


\begin{figure*}
\includegraphics[width=0.77\textwidth,angle=0.0]{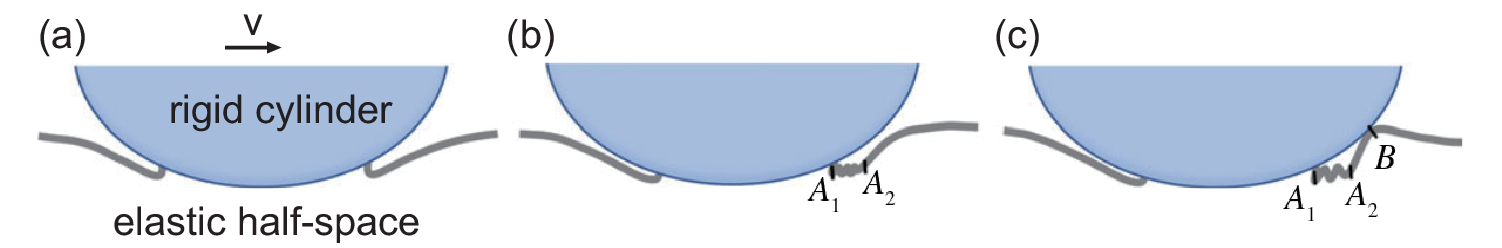}
\caption{\label{Xinstability.eps}
Nucleation of a single Schallamach wave in the cylinder contact geometry. (a) The rigid cylinder moves with a constant
velocity $v$. (b) Buckling results in a pronounced
bulge on the free surface, with a corresponding surface depression between points A1 and A2. (c) The bulge eventually
grows enough to attach to the indenter at point B. Adapted from Ref. \cite{instability}.
}
\end{figure*}

\begin{figure}
\includegraphics[width=0.47\textwidth,angle=0.0]{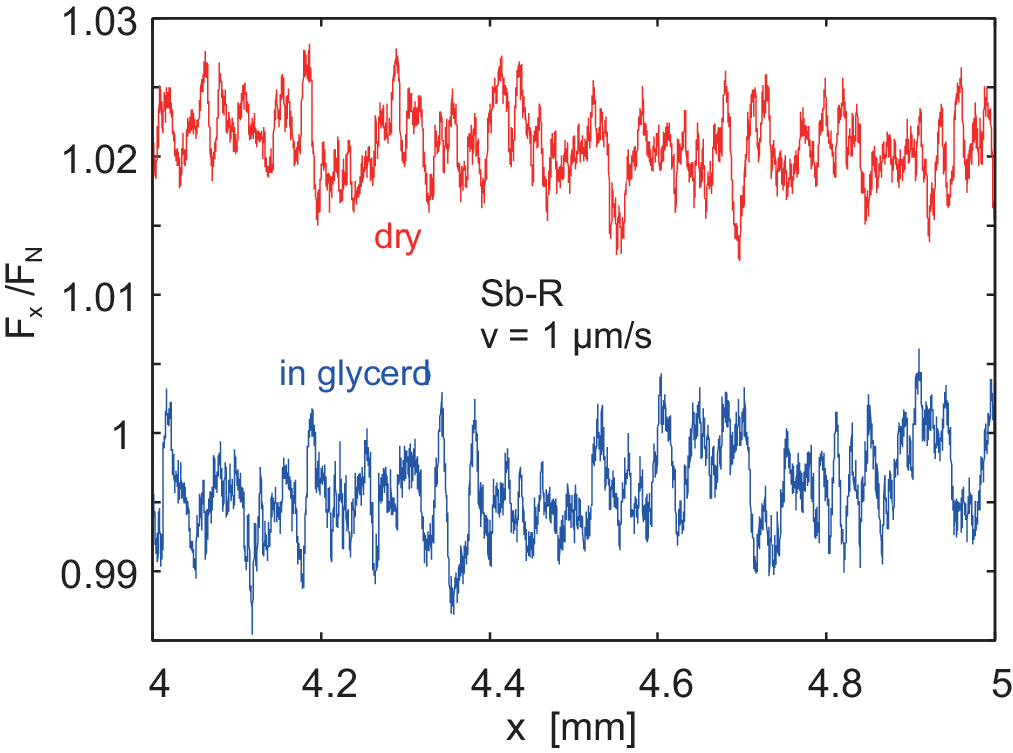}
\caption{\label{1x.2Fx.dryRED.blueWET.rough.eps}
The ratio $\mu = F_x/F_{\rm N}$ as a function of sliding distance for
Sb-R in glycerol (red curve) and in the dry state (blue curve).
The sliding speed is $v=1 \ {\rm \mu m/s}$, and the
normal load is $F_{\rm N}/L = 310 \ {\rm N/m}$. Note the stick-slip oscillations in the friction force.
}
\end{figure}

\begin{figure}
\includegraphics[width=0.47\textwidth,angle=0.0]{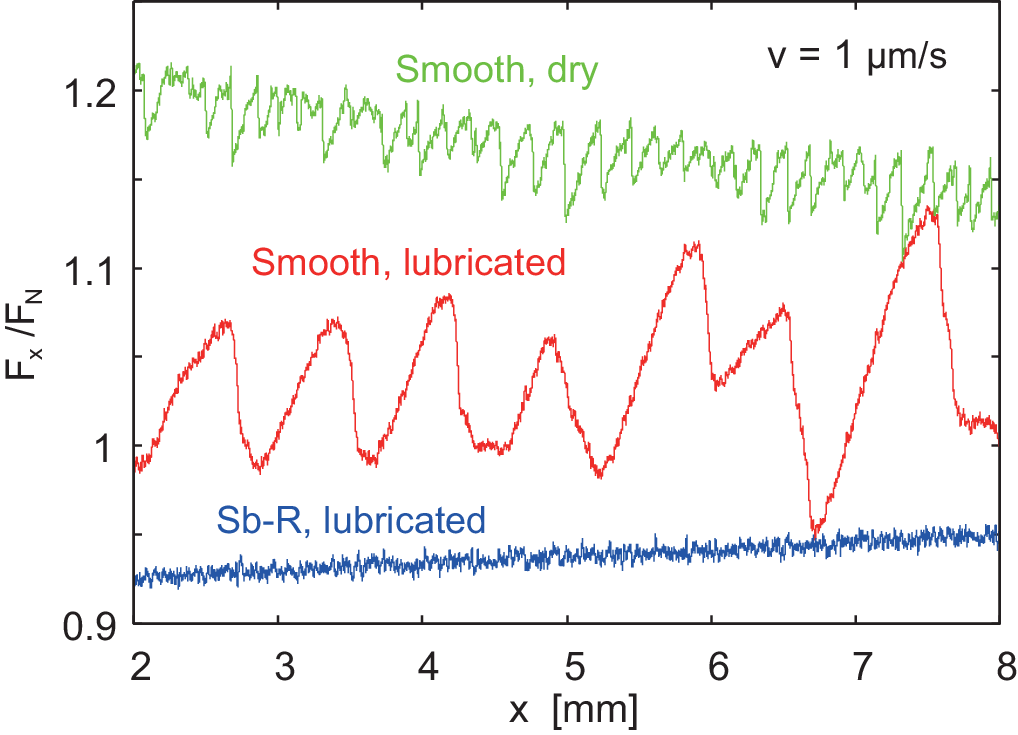}
\caption{\label{1x.2Fx.DRY.WET.WETrough.eps}
The ratio $\mu = F_x/F_{\rm N}$ as a function of sliding distance for the Smooth
surface in glycerol (red curve) and in the dry state (green curve). The blue curve
is for Sb-R in glycerol.
The sliding speed is $v=1 \ {\rm \mu m/s}$, and the normal load is $F_{\rm N}/L = 310 \ {\rm N/m}$.
Note the stick-slip oscillations in the friction force for the Smooth surface.
}
\end{figure}

\vskip 0.3cm
\section{Discussion}

The dry and lubricated friction results together suggest that the effective interfacial shear stress is systematically larger for Sb-R than for Sb-S, while it is much smaller for the Smooth surface.

As discussed in Sec. 5, $\gamma_{\rm eff}(1)=0$ for Sb-S and Sb-R rules out macroscopic adhesion-driven Schallamach waves, but it does not exclude adhesive stick-slip processes at shorter length scales. As discussed in Sec. 6, the larger $\sigma_{\rm f}$ for Sb-R than for Sb-S is consistent with the larger short-wavelength roughness of Sb-R. This roughness can make the real contact area more fragmented, with a larger number of smaller contact spots and contact edges, and may therefore enhance local stick-slip, edge-related dissipation, and pinning in the sliding direction. Here we emphasize that the same trend is also found for the lubricated contacts, indicating that the interfacial shear mechanism identified from the dry-friction study remains relevant in the mixed-lubrication regime.

The Smooth surface is different because it is the only surface for which the macroscopic effective interfacial energy is nonzero. Adhesion can therefore act on the length scale of the nominal contact region and generate macroscopic elastic instabilities, which provides a possible explanation for the much smaller effective $\sigma_{\rm f}$ inferred for this surface.

To see this, it is useful to consider what would happen if no Schallamach waves occurred on the Smooth surface. If sliding took place uniformly over the whole interface, and if the effective interfacial shear stress were similar to that for Sb-S, then the friction coefficient for the Smooth surface would become huge, of order 10, and the macroscopic shear stress would be of order $1 \ {\rm MPa}$. When the shear stress in the nominal contact region becomes of the order of the Young's modulus, elastic instabilities may occur. In this case, detached surface regions may propagate through the contact as Schallamach waves \cite{instability}, see Fig. \ref{Xinstability.eps}. The motion is then not controlled by uniform interfacial slip over the whole nominal contact region, but by instability-mediated motion, which can lead to a much lower effective friction coefficient.

With the present experimental setup, we cannot directly observe what is happening at the buried interface. However, we do have indirect evidence for such instabilities at low sliding speeds, both in dry conditions and in glycerol. If such instabilities did not occur, the theory predicts friction levels much larger than those observed experimentally.

The dry friction traces provide the clearest evidence. Fig. \ref{DRY.eps} shows the ratio $\mu = F_x/F_{\rm N}$ as a function of sliding distance for the Smooth surface and Sb-R in the dry state. The Smooth surface exhibits pronounced stick-slip oscillations, whereas Sb-R does not. The increase in friction force before an abrupt drop may be due to the force required to compress the rubber at the leading edge of the contact, followed by a drop when the detached region enters the contact. When the adhesive bond forms at the leading edge (point B) in Fig. \ref{Xinstability.eps}, the rubber may also exert an attractive force with a component in the forward sliding direction.

The lubricated friction traces show related but modified behavior. Fig. \ref{1x.2Fx.dryRED.blueWET.rough.eps} shows that for Sb-R the fluctuations in glycerol and in the dry state are of similar magnitude, indicating that lubrication does not qualitatively change the sliding mode for the rougher surface. In contrast, Fig. \ref{1x.2Fx.DRY.WET.WETrough.eps} shows that the Smooth surface in glycerol still exhibits large oscillations, much larger than those for Sb-R in glycerol. This is consistent with the fact that adhesion between the rubber and the counter surface can still operate in glycerol for the Smooth surface.

At the same time, the oscillations for the Smooth surface differ between dry and lubricated conditions. In the dry state, they have the typical abrupt stick-slip character, with a nearly linear increase in friction followed by a sudden drop. In glycerol they become more sinus-like. The detailed origin of this difference is not yet clear, but it is likely related to damping by viscous dissipation in the liquid. Finally, the oscillation amplitude is much larger, and the oscillation frequency much lower, at $v=1 \ {\rm \mu m/s}$ than at $v=3 \ {\rm \mu m/s}$. We attribute this to the transition toward a more lubricated interface as the sliding speed increases. This interpretation is consistent with the idea that at low sliding speed the Smooth surface still undergoes an adhesion-controlled sliding mode, whereas at higher sliding speed forced wetting allows a fluid film to enter the nominal contact region and progressively suppresses direct solid-solid contact. Thus, for hydrophobic soft contacts, the Stribeck curve reflects not only fluid flow and load sharing, but also changes in the sliding mechanism of the interface.

\vskip 0.3cm
\section{Summary and conclusions}

In this paper, we have presented an experimental and theoretical study of dry and glycerol-lubricated sliding for PMMA cylinders with different surface roughness sliding on PDMS rubber. The PMMA-PDMS-glycerol system is of particular interest because it represents a hydrophobic lubricated interface, where adhesion may still occur between the solids even in the presence of the lubricant.

The surface roughness and the corresponding contact mechanics are first analyzed. When adhesion is included, the real contact area increases for all surfaces at finite pressure, but the effect is strongest for the Smooth surface and weaker for the two sandblasted surfaces, Sb-S and Sb-R. For the Smooth surface, the macroscopic effective work of adhesion remains nonzero, while for Sb-S and Sb-R it vanishes.

The dry friction measurements provide a useful baseline for the interfacial shear stress. The viscoelastic contribution to friction, calculated using the Persson rubber friction theory and the measured viscoelastic modulus of PDMS, is much smaller than the observed friction for all three surfaces. We therefore conclude that the dry friction is dominated mainly by adhesive molecular interactions at the interface, i.e., by the formation and breaking of adhesive bonds between PMMA and PDMS. By combining the measured dry friction coefficients with the calculated contact areas, we inferred the interfacial shear stress in the real contact area. This analysis shows that the shear stress is largest for Sb-R, smaller for Sb-S, and smallest for the Smooth surface.

For the two sandblasted surfaces, the measured Stribeck curves can be described reasonably well using the mean-field mixed-lubrication theory with a fitted velocity-independent effective interfacial shear stress acting in the area of real contact. 
As in the dry condition, the values of this shear stress are systematically larger for Sb-R than for Sb-S. We attribute this to the larger short-wavelength roughness of Sb-R, which may result in stronger local bonding and/or larger pinning barriers for nanosized rubber patches at the interface.

The Smooth surface behaves in a qualitatively different way. In this case, the mixed-lubrication theory predicts that the boundary-lubrication regime should be reached only at extremely small sliding speeds, whereas the experiments show a friction plateau already at much higher sliding speeds. We attribute this discrepancy to a different sliding state involving adhesion-controlled elastic instabilities, i.e., macroscopic Schallamach-wave-like motion. This interpretation is supported by the measured friction data at low velocities, which show pronounced stick-slip oscillations for the Smooth surface in both dry and lubricated conditions. In glycerol, these oscillations become more sinus-like, which is likely related to viscous damping from the liquid. As the sliding speed increases, forced wetting occurs and a fluid film enters the nominal contact region, which progressively suppresses direct solid-solid contact and the associated instability-mediated sliding mode.

One of the most important findings of this study is that the physics of the Stribeck curve for soft hydrophobic contacts cannot always be understood from the standard mixed-lubrication theory alone. For rough surfaces, the standard mixed-lubrication picture works reasonably well when combined with an effective interfacial shear stress. For sufficiently smooth and adhesive surfaces, however, adhesion changes not only the real contact area but also the sliding mode itself. In this case, the Stribeck curve reflects a competition of interfacial adhesion, instability-mediated sliding, and forced wetting.

\vskip 0.3cm
\section{Appendix A: numerical results for flow and friction factors}

Here we present the fluid flow and friction factors for all three surfaces studied in
this paper. The analytical expressions for these factors were given in Ref. \cite{PS3} and formally
result from integrating out the roughness in the determination of fluid flow.

\begin{figure}
\includegraphics[width=0.47\textwidth,angle=0.0]{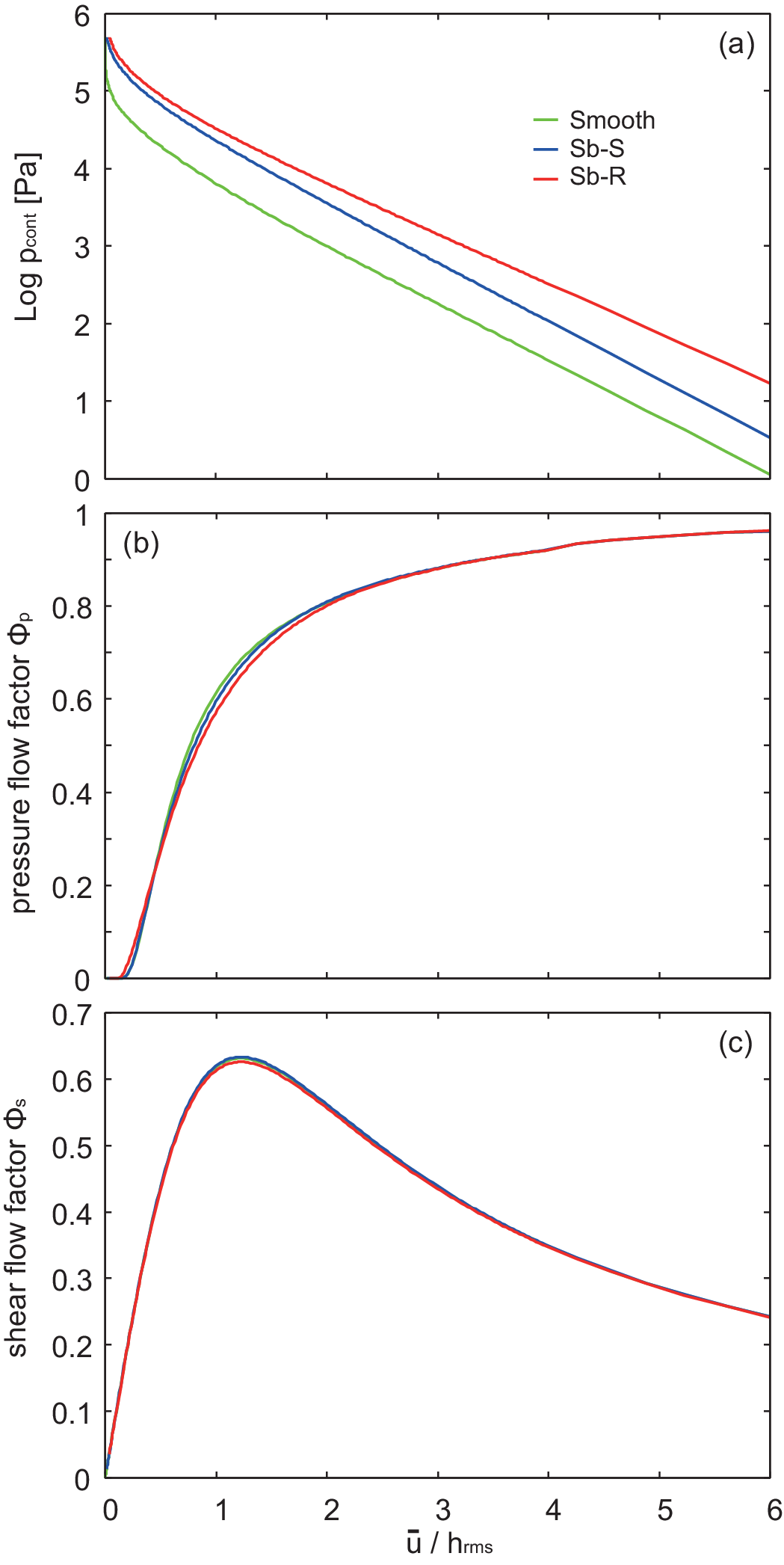}
\caption{\label{1separation.2fluidfowfactors.eps}
The logarithm of the contact pressure $\bar p_{\rm cont}$ and the fluid flow factors $\phi_{\rm p}$ and $\phi_{\rm s}$
as functions of the surface separation normalized by the
rms roughness amplitude $h_{\rm rms}$. For the Smooth surface (green line, $h_{\rm rms} = 1.372 \ {\rm \mu m}$),
Sb-S (blue line, $h_{\rm rms} = 2.798 \ {\rm \mu m}$), and Sb-R
(red line, $h_{\rm rms} = 15.76 \ {\rm \mu m}$).
}
\end{figure}

Fig. \ref{1separation.2fluidfowfactors.eps}
shows the logarithm of the contact pressure $\bar p_{\rm cont}$ and the fluid flow factors $\phi_{\rm p}$ and $\phi_{\rm s}$
as functions of the surface separation $\bar u$ normalized by the
rms roughness amplitude $h_{\rm rms}$. The green curves are for the Smooth surface with $h_{\rm rms} = 1.372 \ {\rm \mu m}$,
the blue lines are for Sb-S, with $h_{\rm rms} = 2.798 \ {\rm \mu m}$, and the red lines are for
Sb-R, with $h_{\rm rms} = 15.76 \ {\rm \mu m}$. Note that the pressure flow factor vanishes when the
contact area percolates, which occurs when $\bar u \approx 0.2 h_{\rm rms}$ for all three surfaces.
The general shapes of the flow factor curves are similar to those obtained by Patir and Cheng \cite{Pati1,Pati2}.

\begin{figure}
\includegraphics[width=0.47\textwidth,angle=0.0]{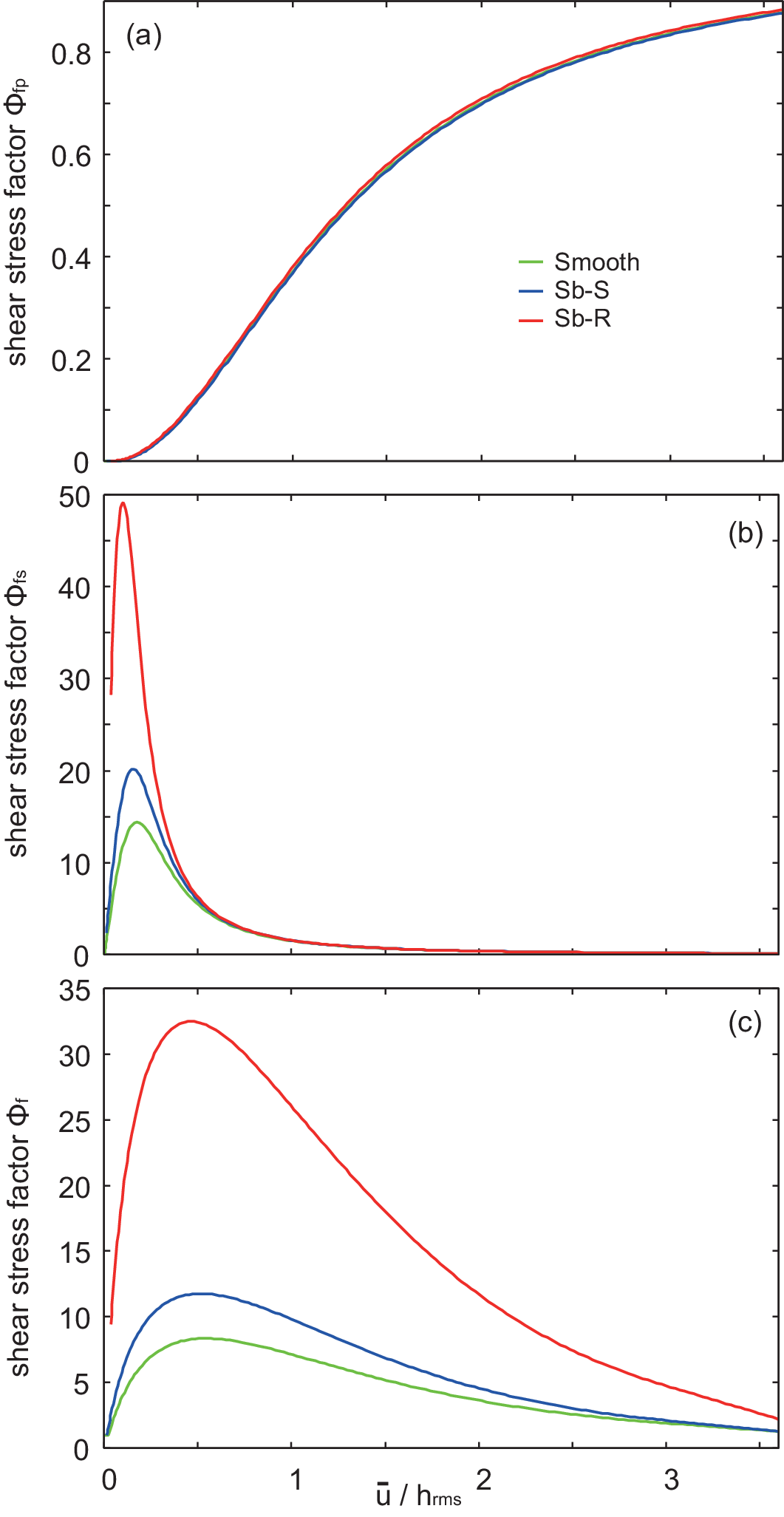}
\caption{\label{1separation.2frictionfactors.eps}
The friction factors $\phi_{\rm fp}$, $\phi_{\rm fs}$, and $\phi_{\rm f}$ as functions of the surface separation normalized by the
rms roughness amplitude $h_{\rm rms}$. For the Smooth surface (green line, $h_{\rm rms} = 1.372 \ {\rm \mu m}$),
Sb-S (blue line, $h_{\rm rms} = 2.798 \ {\rm \mu m}$), and Sb-R
(red line, $h_{\rm rms} = 15.76 \ {\rm \mu m}$).
}
\end{figure}

\begin{figure}
\includegraphics[width=0.47\textwidth,angle=0.0]{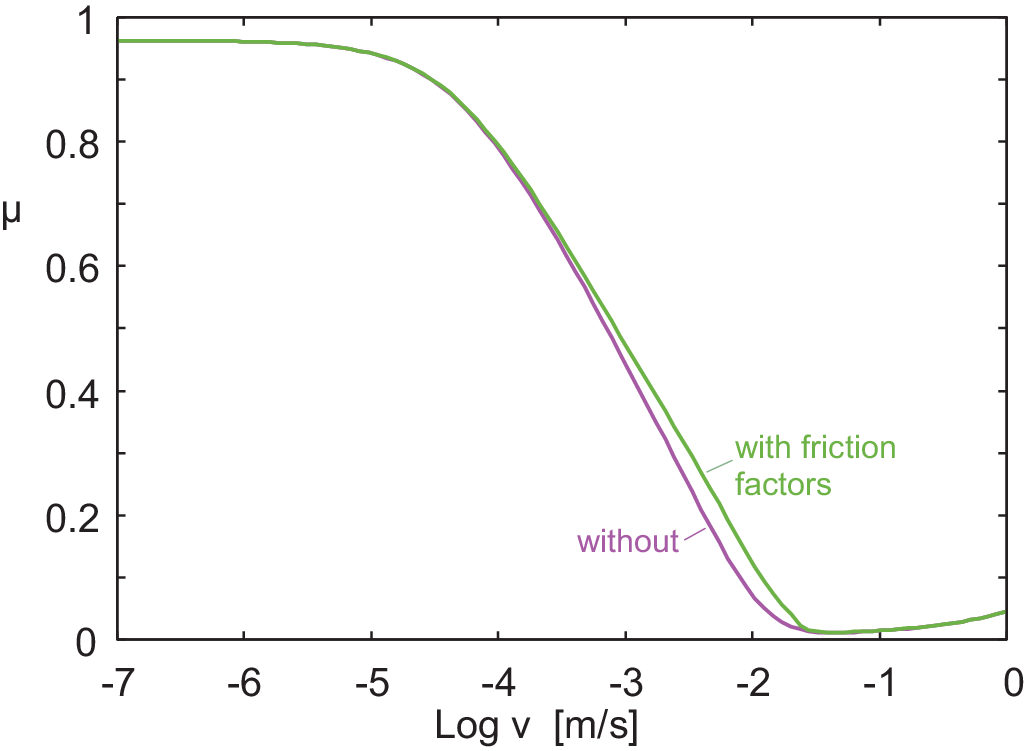}
\caption{\label{1logv.2mu.with.without.frictionfactors.rough3.eps}
The friction coefficient as a function of the logarithm of sliding speed obtained with and without friction factors for Sb-R. The calculations use $F_{\rm N}/L = 2110 \ {\rm N/m}$, $\eta = 1.4 \ {\rm Pa\,s}$, $R = 7 \ {\rm cm}$, and $\sigma_{\rm f} = 0.98 \ {\rm MPa}$.
For Sb-S, the calculations with and without the friction factors are nearly identical and are therefore not shown.
}
\end{figure}

Fig. \ref{1separation.2frictionfactors.eps}
shows the friction factors $\phi_{\rm fp}$, $\phi_{\rm fs}$, and $\phi_{\rm f}$ as functions of the surface separation normalized by the
rms roughness amplitude $h_{\rm rms}$. For the present systems and conditions, the friction factors have a negligible influence on the
Stribeck curve for the Smooth surface and for Sb-S, while they result in a small
modification of the Stribeck curve for Sb-R. This is illustrated in Fig. \ref{1logv.2mu.with.without.frictionfactors.rough3.eps}
for Sb-R with the normal load $F_{\rm N}/L = 2110 \ {\rm N/m}$. However, we showed in an earlier study that
for other surfaces the friction factors, in particular $\phi_{\rm f}$, can have a large influence on the Stribeck curve and can generate
a local maximum in the $\mu (v)$ curve close to the velocity where the $\mu (v)$ curve without the friction factors
would have a minimum \cite{PS3}. Such a maximum was observed in the experiments reported in Ref. \cite{Din}.

\vskip 0.3cm
\section{Appendix B: numerical results for the fluid contact pressure and interfacial separation}

Here we present results for the average surface separation and the contact pressure for Sb-S and Sb-R.
We assume the normal load $F_{\rm N}/L = 310 \ {\rm N/m}$ and consider a few different sliding speeds.

For a perfectly smooth surface at vanishing sliding speed, the contact pressure is Hertzian, with a
maximum of $0.0531 \ {\rm MPa}$ and a contact width of $\approx 7.43 \ {\rm mm}$. For the rough
surfaces, the width of the nominal contact region increases due to contact with ``high'' asperities outside the Hertzian
contact, and the pressure distribution changes continuously from Hertzian-like to Gaussian-like \cite{Johnson,Gauss} with
increasing roughness amplitude.

\begin{figure}
\includegraphics[width=0.47\textwidth,angle=0.0]{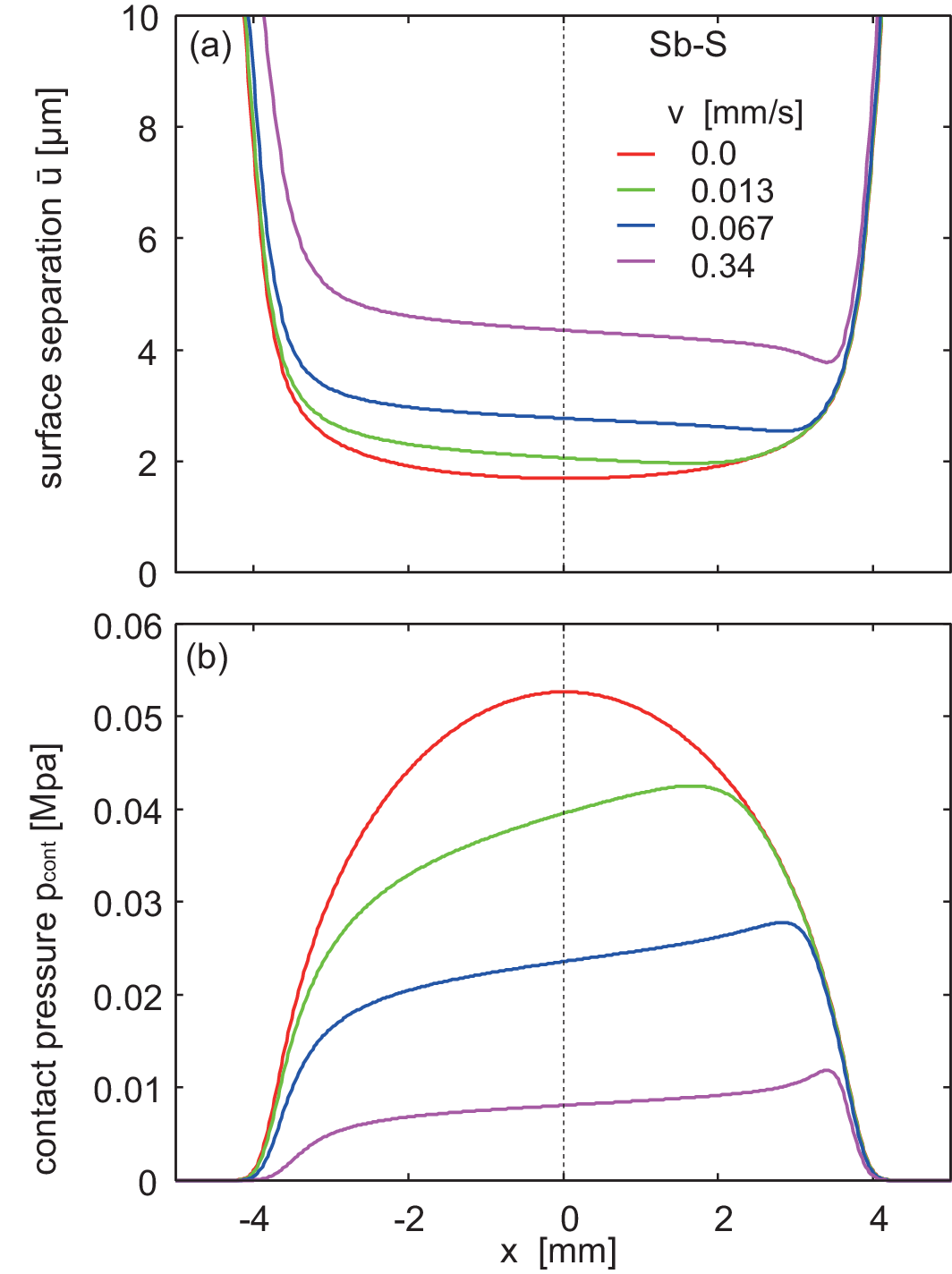}
\caption{\label{1x.2separation.pcont.smooth0.a0.b30.c40.d50.eps}
The surface separation $\bar u$ and the contact pressure $p_{\rm cont}$ as functions of the lateral coordinate
$x$ for Sb-S and the load $F_{\rm N}/L = 310 \ {\rm N/m}$
at different sliding speeds $v$.
}
\end{figure}

Fig. \ref{1x.2separation.pcont.smooth0.a0.b30.c40.d50.eps}
shows the surface separation $\bar u$ and the contact pressure $p_{\rm cont}$ as functions of the lateral coordinate
$x$ for Sb-S. For $v=0$ the maximum contact pressure and the width of the pressure distribution are
$\approx 0.0526 \ {\rm MPa}$ and $\approx 8.34 \ {\rm mm}$, respectively.

\begin{figure}
\includegraphics[width=0.47\textwidth,angle=0.0]{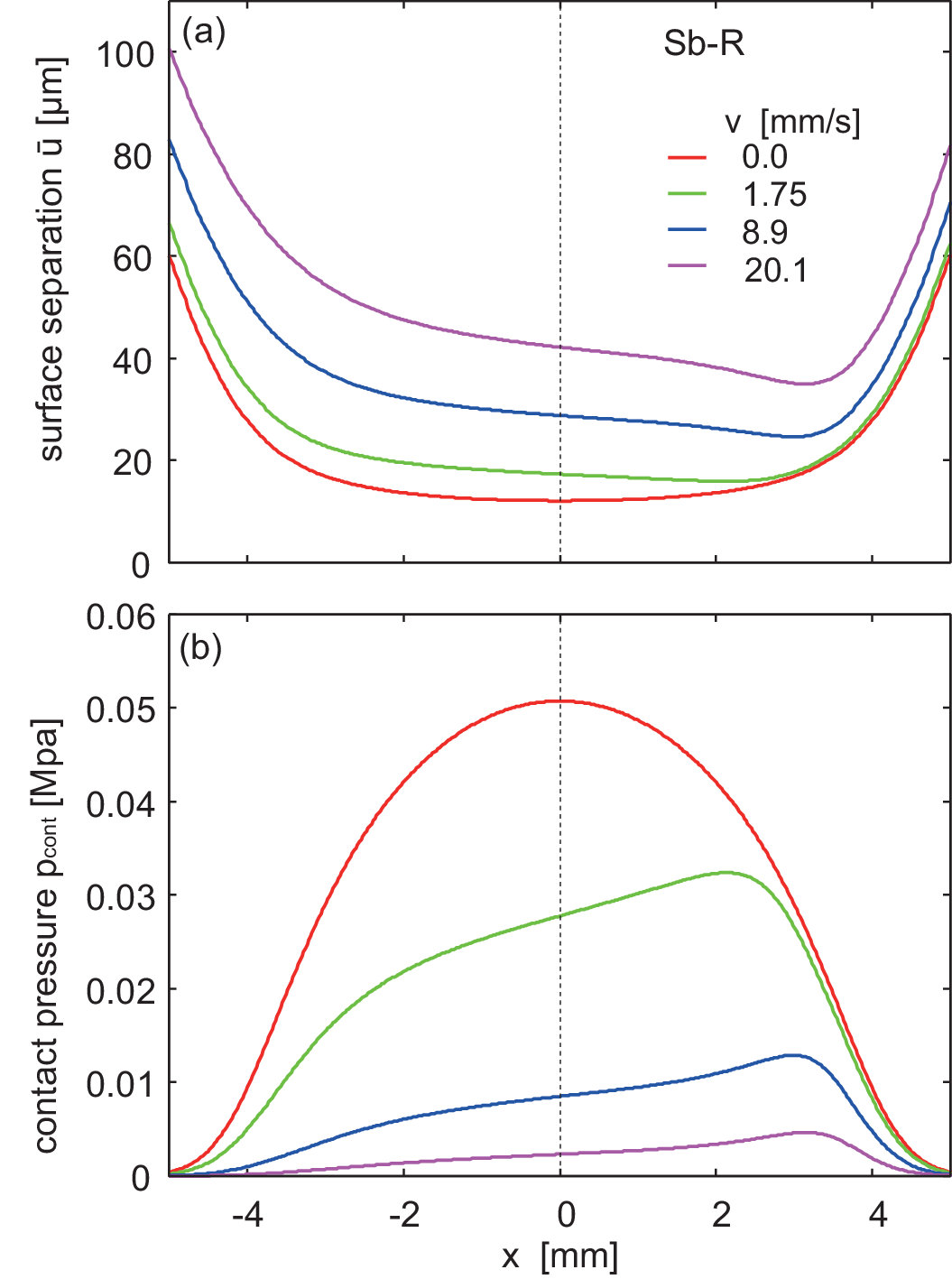}
\caption{\label{1x.2separation.pcont.rough0.a0.b60.c70.d75.eps}
The surface separation $\bar u$ and the contact pressure $p_{\rm cont}$ as functions of the lateral coordinate
$x$ for Sb-R and the load $F_{\rm N}/L = 310 \ {\rm N/m}$ at different sliding speeds $v$.
}
\end{figure}

Fig. \ref{1x.2separation.pcont.rough0.a0.b60.c70.d75.eps}
shows the surface separation $\bar u$ and the contact pressure $p_{\rm cont}$ as functions of the lateral coordinate
$x$ for Sb-R. For $v=0$ the maximum contact pressure and the width of the pressure distribution are
$\approx 0.0507 \ {\rm MPa}$ and $\approx 10.64 \ {\rm mm}$, respectively.

For both surfaces, the interfacial separation at ``high'' sliding speeds has the
characteristic form expected from elastohydrodynamics, with the minimum separation close to the exit of the nominal
contact region. At the point where the separation is smallest, the contact pressure is, as expected, maximal.

\begin{figure}
\includegraphics[width=0.47\textwidth,angle=0.0]{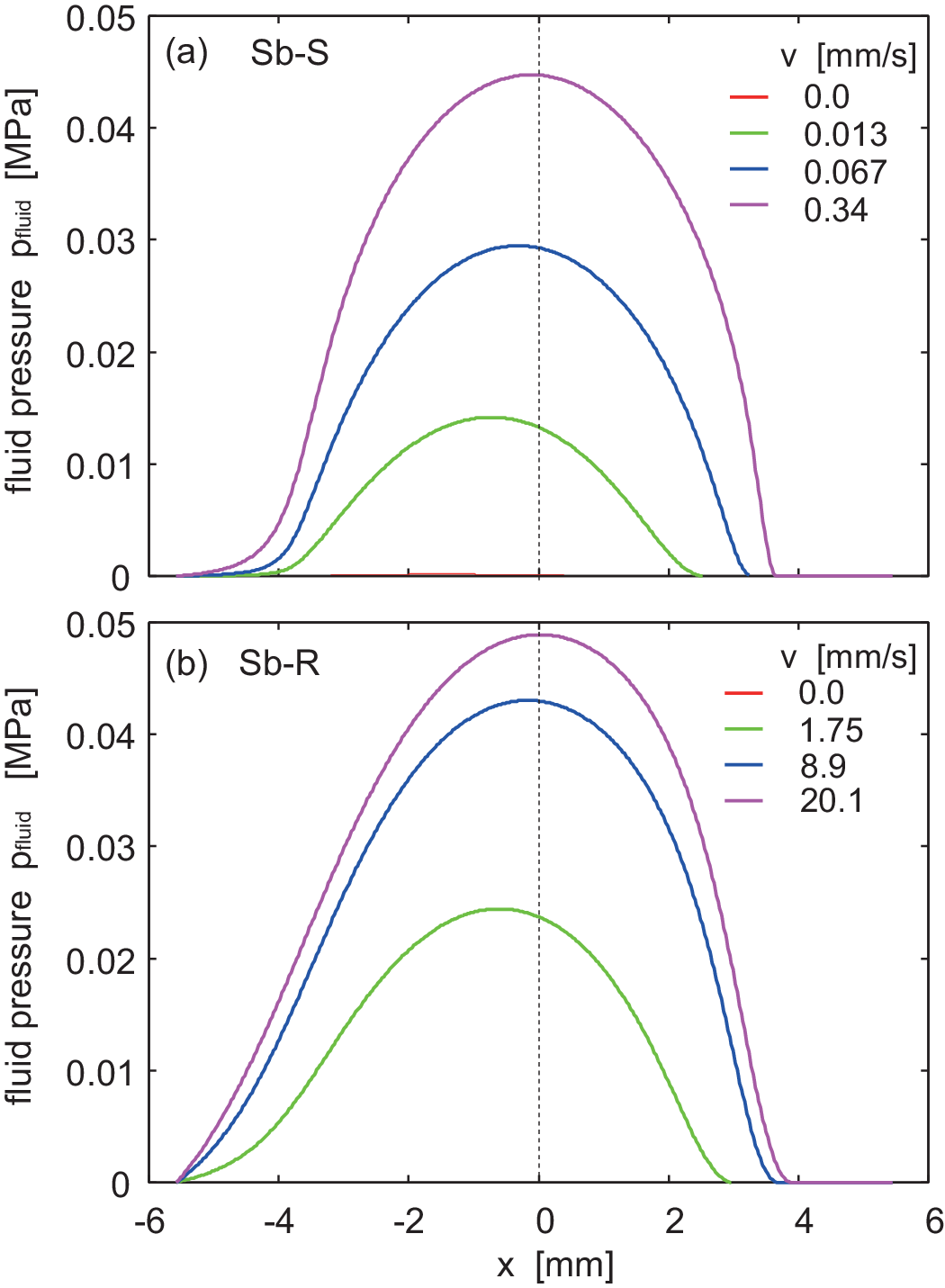}
\caption{\label{1fluidPsmoothPrough.eps}
The fluid pressure $\bar p_{\rm fluid}$ as a function of the lateral coordinate
$x$ for Sb-S and Sb-R and the load $F_{\rm N}/L = 310 \ {\rm N/m}$ at different sliding speeds $v$.
}
\end{figure}

Fig. \ref{1fluidPsmoothPrough.eps} shows the
fluid pressure $\bar p_{\rm fluid}$ as a function of the lateral coordinate
$x$ for Sb-S and Sb-R at different sliding speeds $v$.
The fluid pressure profile is ``tilted'' toward the entrance side, and there are no negative pressures anywhere at the interface.
We note that in the present calculations we include cavitation. For nearly rigid surfaces, cavitation is needed in order for the
fluid-induced normal (load-bearing) force to be nonzero, because the pressure profile 
for rigid solids without cavitation is antisymmetric around the contact midpoint,
and the fluid pressure integrated over $x$ vanishes. 
However, for elastically soft solids like the PDMS rubber used here, the friction force calculated with and without
cavitation is nearly the same (see Ref. \cite{PS1}).

\vskip 0.3cm
\section{Appendix C: Calculation of friction coefficients with error bars}

Here we describe how the friction coefficients shown in the main text were obtained from the raw data. We also present the corresponding standard-deviation error bars.

The friction coefficient was first calculated as $\mu = F_x/F_{\rm N}$ for each data point. For each imposed sliding speed, the corresponding constant-velocity segment was identified from the velocity signal. The first and final $10\%$ of each segment were excluded to avoid the initial transient and the transition to the next velocity.

\begin{figure}
\centering
\includegraphics[width=0.47\textwidth,angle=0.0]{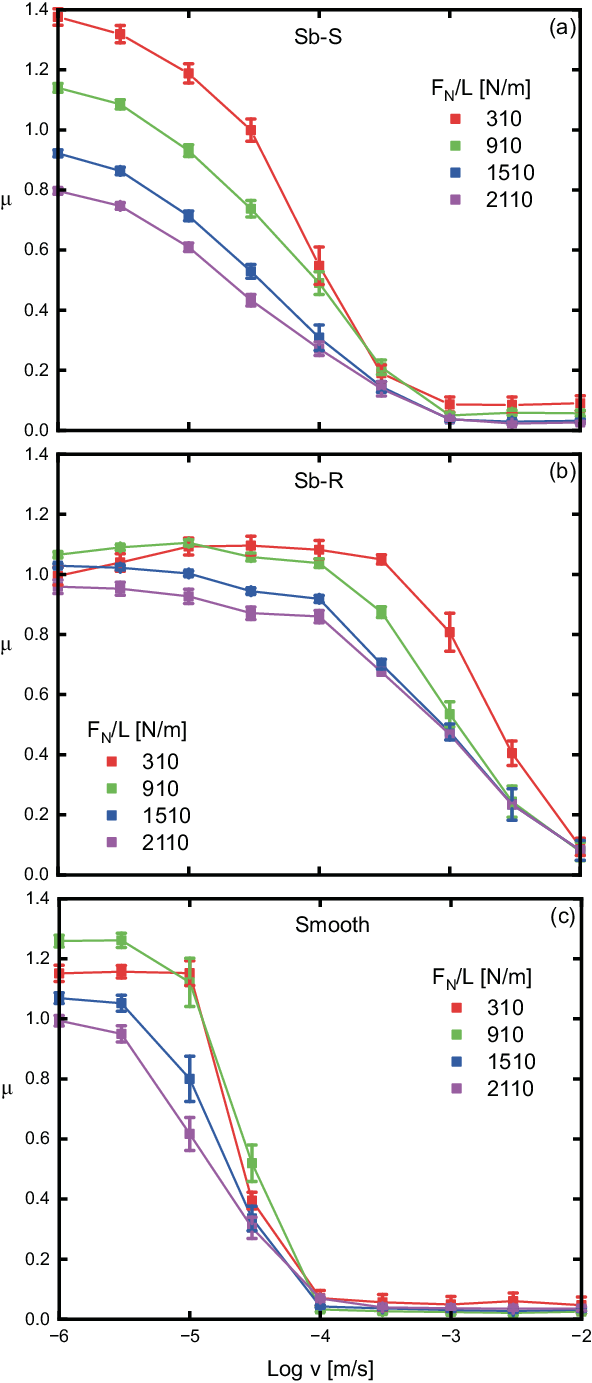}
\caption{\label{friction.errorbars.eps}
Friction coefficient as a function of the logarithm of sliding speed under glycerol-lubricated conditions for
(a) Sb-S, (b) Sb-R, and (c) the Smooth surface.
The error bars indicate the standard deviation of the friction coefficient within the selected region.
The lines connecting the data points are only guides to the eye and do not represent theoretical fits.
}
\end{figure}

Fig. \ref{friction.errorbars.eps} shows the same experimental data as in the main text, but with error bars indicating the standard deviation of the friction coefficient within the selected region.

\end{document}